\documentclass[
reprint,
superscriptaddress,
nofootinbib,
amsmath,amssymb,
aps,
prd,
floatfix]{revtex4-2}
 
\usepackage{xcolor}
\usepackage{graphicx}% Include figure files
\usepackage{dcolumn}% Align table columns on decimal point
\usepackage{bm}% bold math
\usepackage[pdftex]{hyperref} % Extensive support for hypertext in LATEX
\usepackage[mathlines]{lineno}% Enable numbering of text and display math

\usepackage{svg}

\usepackage{adjustbox}
\usepackage{multirow}
\usepackage{lineno}

\newcommand{\umu}{$\lvert U_{\mu4}\rvert^2$}
\newcommand{\umusq}{\lvert U_{\mu4}\rvert^2}

\newcommand{\mumix}{\lvert U_{\mu 4}\rvert^2}

\newcommand{\thetaeffmix}{\theta_{\textrm{eff}}^2}
\newcommand{\thetamix}{\theta^2}
\def\mhps {{\ensuremath{m_{\mathrm{HPS}}}}}
\def\mhnl {{\ensuremath{m_{\mathrm{HNL}}}}}
\def\mllp {{\ensuremath{m_{\mathrm{LLP}}}}}
\def\mllpsq {{\ensuremath{m^2_{\mathrm{LLP}}}}}
\begin{document}

\widetext
\leftline{FERMILAB-PUB-22-507-ND}

%\centerline{\em INTERNAL DOCUMENT -- NOT FOR PUBLIC DISTRIBUTION}

\title{Search for long-lived heavy neutral leptons and Higgs portal scalars decaying in the MicroBooNE detector}

% List of institutions in command form:
\newcommand{\Bern}{Universit{\"a}t Bern, Bern CH-3012, Switzerland}
\newcommand{\BNL}{Brookhaven National Laboratory (BNL), Upton, NY, 11973, USA}
\newcommand{\UCSB}{University of California, Santa Barbara, CA, 93106, USA}
\newcommand{\Cambridge}{University of Cambridge, Cambridge CB3 0HE, United Kingdom}
\newcommand{\CIEMAT}{Centro de Investigaciones Energ\'{e}ticas, Medioambientales y Tecnol\'{o}gicas (CIEMAT), Madrid E-28040, Spain}
\newcommand{\Chicago}{University of Chicago, Chicago, IL, 60637, USA}
\newcommand{\Cincinnati}{University of Cincinnati, Cincinnati, OH, 45221, USA}
\newcommand{\CSU}{Colorado State University, Fort Collins, CO, 80523, USA}
\newcommand{\Columbia}{Columbia University, New York, NY, 10027, USA}
\newcommand{\Edinburgh}{University of Edinburgh, Edinburgh EH9 3FD, United Kingdom}
\newcommand{\FNAL}{Fermi National Accelerator Laboratory (FNAL), Batavia, IL 60510, USA}
\newcommand{\Granada}{Universidad de Granada, Granada E-18071, Spain}
\newcommand{\Harvard}{Harvard University, Cambridge, MA 02138, USA}
\newcommand{\IIT}{Illinois Institute of Technology (IIT), Chicago, IL 60616, USA}
\newcommand{\KSU}{Kansas State University (KSU), Manhattan, KS, 66506, USA}
\newcommand{\Lancaster}{Lancaster University, Lancaster LA1 4YW, United Kingdom}
\newcommand{\LANL}{Los Alamos National Laboratory (LANL), Los Alamos, NM, 87545, USA}
\newcommand{\Louisiana}{Louisiana State University, Baton Rouge, LA, 70803, USA}
\newcommand{\Manchester}{The University of Manchester, Manchester M13 9PL, United Kingdom}
\newcommand{\MIT}{Massachusetts Institute of Technology (MIT), Cambridge, MA, 02139, USA}
\newcommand{\Michigan}{University of Michigan, Ann Arbor, MI, 48109, USA}
\newcommand{\Minnesota}{University of Minnesota, Minneapolis, MN, 55455, USA}
\newcommand{\NMSU}{New Mexico State University (NMSU), Las Cruces, NM, 88003, USA}
\newcommand{\Oxford}{University of Oxford, Oxford OX1 3RH, United Kingdom}
\newcommand{\Pitt}{University of Pittsburgh, Pittsburgh, PA, 15260, USA}
\newcommand{\Rutgers}{Rutgers University, Piscataway, NJ, 08854, USA}
\newcommand{\SLAC}{SLAC National Accelerator Laboratory, Menlo Park, CA, 94025, USA}
\newcommand{\SDSMT}{South Dakota School of Mines and Technology (SDSMT), Rapid City, SD, 57701, USA}
\newcommand{\Maine}{University of Southern Maine, Portland, ME, 04104, USA}
\newcommand{\Syracuse}{Syracuse University, Syracuse, NY, 13244, USA}
\newcommand{\TelAviv}{Tel Aviv University, Tel Aviv, Israel, 69978}
\newcommand{\Tennessee}{University of Tennessee, Knoxville, TN, 37996, USA}
\newcommand{\UTA}{University of Texas, Arlington, TX, 76019, USA}
\newcommand{\Tufts}{Tufts University, Medford, MA, 02155, USA}
\newcommand{\VTech}{Center for Neutrino Physics, Virginia Tech, Blacksburg, VA, 24061, USA}
\newcommand{\Warwick}{University of Warwick, Coventry CV4 7AL, United Kingdom}
\newcommand{\Yale}{Wright Laboratory, Department of Physics, Yale University, New Haven, CT, 06520, USA}
%%\newcommand{\listerThanks}{Now at University of Wisconsin, Madison}

% So that institutions appear in alphabetical order:
\affiliation{\Bern}
\affiliation{\BNL}
\affiliation{\UCSB}
\affiliation{\Cambridge}
\affiliation{\CIEMAT}
\affiliation{\Chicago}
\affiliation{\Cincinnati}
\affiliation{\CSU}
\affiliation{\Columbia}
\affiliation{\Edinburgh}
\affiliation{\FNAL}
\affiliation{\Granada}
\affiliation{\Harvard}
\affiliation{\IIT}
\affiliation{\KSU}
\affiliation{\Lancaster}
\affiliation{\LANL}
\affiliation{\Louisiana}
\affiliation{\Manchester}
\affiliation{\MIT}
\affiliation{\Michigan}
\affiliation{\Minnesota}
\affiliation{\NMSU}
\affiliation{\Oxford}
\affiliation{\Pitt}
\affiliation{\Rutgers}
\affiliation{\SLAC}
\affiliation{\SDSMT}
\affiliation{\Maine}
\affiliation{\Syracuse}
\affiliation{\TelAviv}
\affiliation{\Tennessee}
\affiliation{\UTA}
\affiliation{\Tufts}
\affiliation{\VTech}
\affiliation{\Warwick}
\affiliation{\Yale}

% Authors in alphabetical order
\author{P.~Abratenko} \affiliation{\Tufts}
\author{J.~Anthony} \affiliation{\Cambridge}
\author{L.~Arellano} \affiliation{\Manchester}
\author{J.~Asaadi} \affiliation{\UTA}
\author{A.~Ashkenazi}\affiliation{\TelAviv}
\author{S.~Balasubramanian}\affiliation{\FNAL}
\author{B.~Baller} \affiliation{\FNAL}
\author{C.~Barnes} \affiliation{\Michigan}
\author{G.~Barr} \affiliation{\Oxford}
\author{J.~Barrow} \affiliation{\MIT}\affiliation{\TelAviv}
% ...

\collaboration{The MicroBooNE Collaboration}
\thanks{microboone\_info@fnal.gov}\noaffiliation

\date{\today}

\begin{abstract}
We present a search for long-lived Higgs portal scalars (HPS) and heavy neutral leptons (HNL) decaying in the MicroBooNE liquid-argon time projection chamber.  The measurement is performed using data collected synchronously with the NuMI neutrino beam from Fermilab's Main Injector with a total exposure corresponding to $7.01 \times 10^{20}$ protons on target.
We set upper limits at the $90\%$ confidence level on the mixing parameter $\lvert U_{\mu 4}\rvert^2$ ranging from $\lvert U_{\mu 4}\rvert^2<12.9\times 10^{-8}$ for Majorana HNLs with a mass of $m_{\rm HNL}=246$~MeV to $\lvert U_{\mu 4}\rvert^2<0.92 \times 10^{-8}$ for $\mhnl=385$~MeV, assuming $\lvert U_{e 4}\rvert^2 = \lvert U_{\tau 4}\rvert^2 = 0$ and HNL decays into $\mu^\pm\pi^\mp$ pairs.
These limits on $\lvert U_{\mu 4}\rvert^2$ represent an order of magnitude improvement in sensitivity compared to the previous MicroBooNE result.
We also constrain the scalar-Higgs mixing angle $\theta$ by searching for HPS decays into $\mu^+\mu^-$ final states, excluding a contour in the parameter space with lower bounds of 
$\theta^2<31.3 \times 10^{-9}$ for $m_{\rm HPS}=212$~GeV 
and
$\theta^2<1.09 \times 10^{-9}$ for $m_{\rm HPS}=275$~GeV.
These are the first constraints on the scalar-Higgs mixing angle $\theta$ from a dedicated experimental search in this mass range.
\end{abstract}

\maketitle

%% SECTIONS

\section{Introduction}

The MicroBooNE detector \cite{Acciarri:2016smi} began collecting data in 2015, making it the first fully operational detector of the three liquid-argon time projection chambers comprising the Short-Baseline Neutrino (SBN) program~\cite{Machado:2019oxb} at Fermilab. The MicroBooNE detector was exposed to both the booster neutrino beam (BNB)~\cite{AguilarArevalo:2008yp} and
the neutrino beam from the main injector (NuMI)~\cite{ADAMSON2016279}. 

We can use the NuMI beam to study beyond-the-Standard Model (BSM) phenomena such as the production and decay of heavy neutral leptons (HNL) or Higgs portal scalars (HPS), jointly referred to as long-lived particles (LLPs). In addition, we have also
 used it to measure electron neutrino cross sections on argon~\cite{MicroBooNENuMInuexsec1,MicroBooNENuMInuexsec2}.

\begin{figure}[htb]
\includegraphics[width=0.48\textwidth]{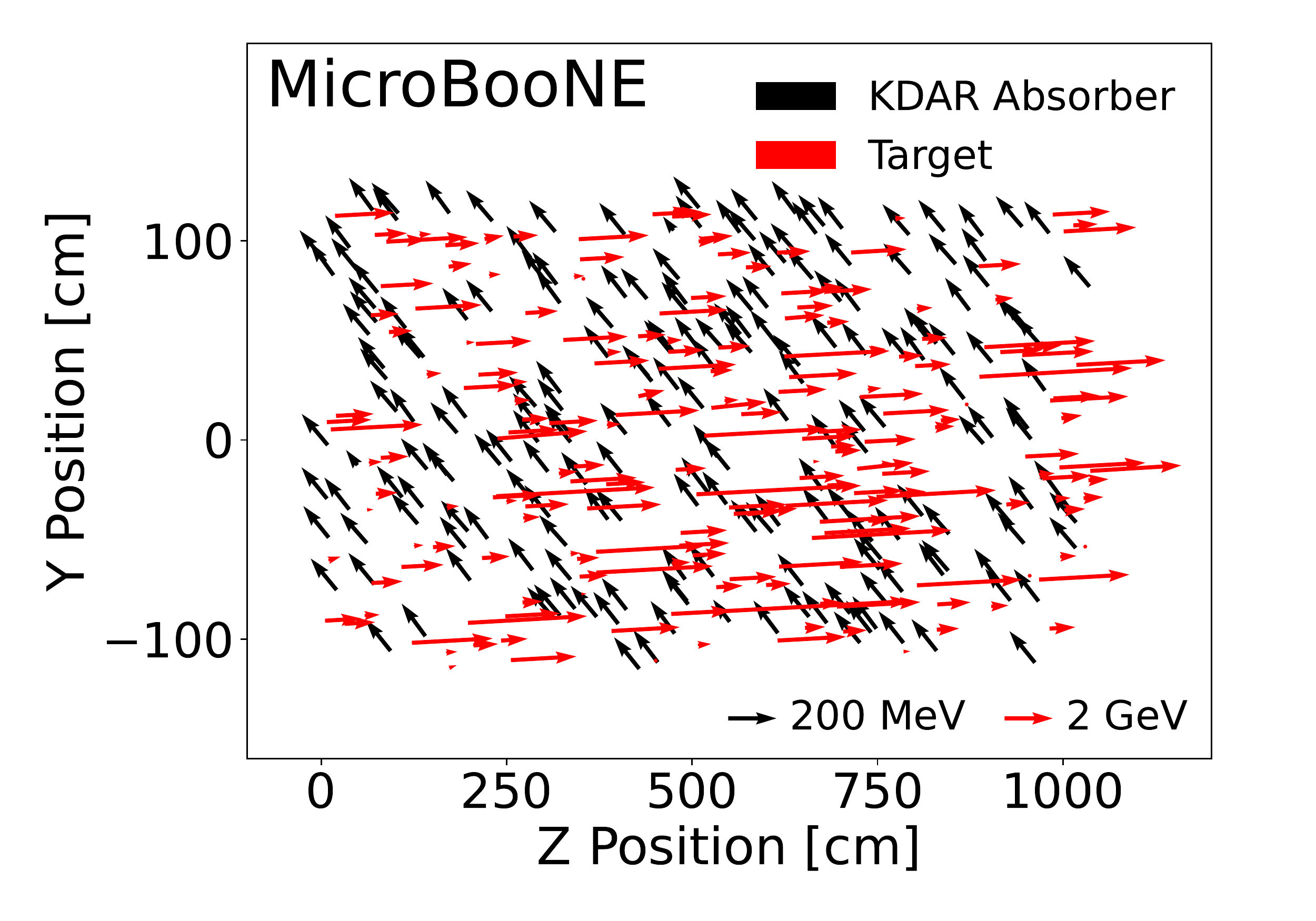}
\caption{Momentum vectors of simulated neutrino events from the NuMI target (red) and of HPS decays (black), where the HPS originated from kaons decaying at rest (KDAR) in the hadron absorber. The vectors are shown in the $y-z$ plane of the MicroBooNE coordinate system within the detector's active volume, where $y$ points upward and $z$ points in the nominal BNB beam direction. The vectors start at the vertex location and their lengths are proportional to the magnitude of the momentum of the neutrino or the HPS.
Different momentum scales are used for the display.} 
\label{fig:HNL_decay}
\end{figure}

\begin{figure*}[hbt]
  \centering
  \includegraphics[width=0.45\textwidth]{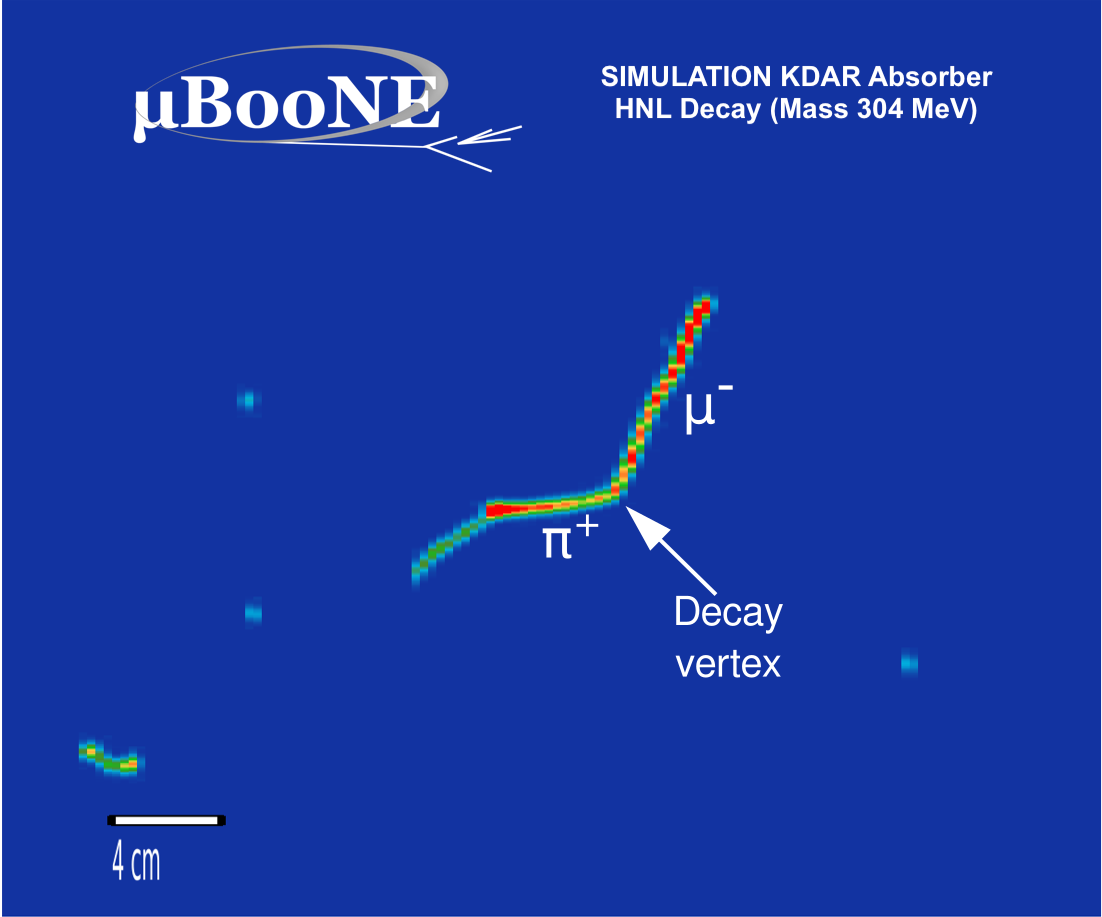}
  \includegraphics[width=0.45\textwidth]{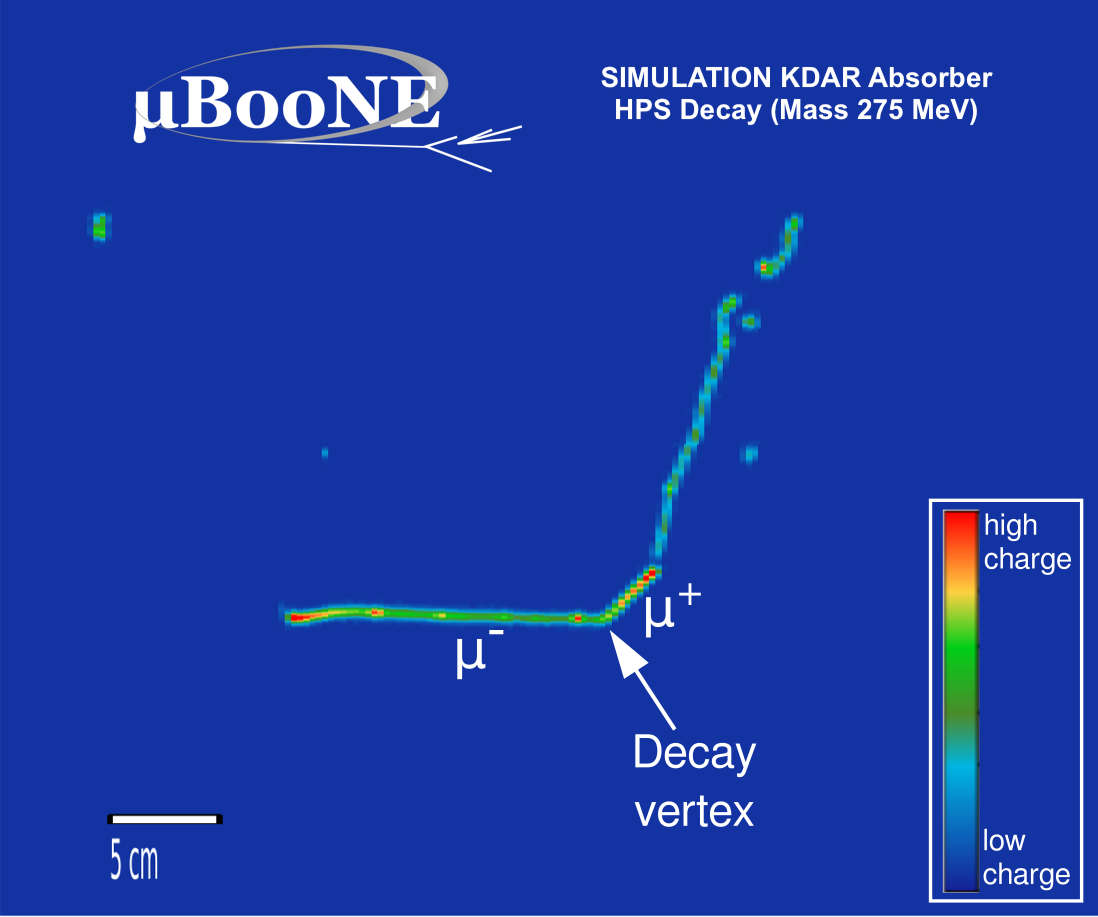}
  \caption{Displays of two simulated signal events, where the LLPs approach from the bottom right of the image and then decay in the detector. The left display shows an HNL of mass $\mhnl=304$~MeV decaying into a muon (track pointing up and right) and a charged pion (track pointing left) that itself subsequently decays into a muon. The right display shows an HPS of mass $\mhps=275$~MeV decaying into a $\mu^-$ (the long track going left) and a $\mu^+$ (the shorter track) which quickly decays to a Michel positron.
  The horizontal direction represents the wires on the collection plane and the
vertical direction represents the electron drift time. Colors
represent the amount of charge deposited on wires.}
  \label{fig:sigeventdisplays}
\end{figure*}
In this paper, we present searches for both types of LLPs originating from kaons decaying at rest in the NuMI hadron absorber, which is located downstream of the MicroBooNE detector at the end of the NuMI decay pipe.
The LLP would travel 
$\approx 104$~m to the MicroBooNE detector where it can be detected through its decay. 
 LLPs produced in the absorber would approach the detector in almost the opposite direction to the vast majority of neutrinos, which originate from near the NuMI beam target (Fig.~\ref{fig:HNL_decay}). 

Event displays of simulated HNL and HPS decays in the MicroBooNE detector are
shown in Fig.\ref{fig:sigeventdisplays}.
The signal topology is characterized by exactly two tracks emerging from a common vertex.
Since we search for LLPs produced by two-body decays of kaons at rest, these LLPs have a fixed energy for a given mass.  These two properties, the direction and energy of the signal LLPs, help to discriminate them from neutrino and cosmic-ray induced background processes. In addition, the kinematics and topologies of HPS decays to $\mu\mu$ pairs and HNL decays to $\mu\pi$ pairs are similar, which allows us to develop a single signal analysis strategy~\cite{Goodwin:2022jej}.

The MicroBooNE collaboration has published upper limits on the production of HNLs decaying to $\mu \pi$ pairs for an exposure of $2.0 \times 10^{20}$ protons on target (POT) from the BNB, using a dedicated trigger configured to detect HNL decays that occur after the neutrino spill reaches the detector. That search yielded upper limits at the $90\%$ confidence level (CL) on the element $\umusq$ of the extended 
Pontecorvo–Maki–Nakagawa–Sakata (PMNS) mixing matrix $\umusq$ for Dirac and Majorana HNLs
in the HNL mass range $260\le \mhnl\le 385$~MeV and assuming $\lvert U_{e 4}\rvert^2 = \lvert U_{\tau 4}\rvert^2 = 0$~\cite{Abratenko:2019kez}.
Several other collaborations have also set limits on
$\umusq$~\cite{Abe:2019kgx,Vaitaitis:1999wq,Bernardi:1985ny,Bernardi:1987ek,CortinaGil:2017mqf,Artamonov:2014urb,Asano:1981he,Hayano:1982wu}.
Also using a liquid-argon detector exposed to the NuMI beam, 
the ArgoNeuT collaboration has published a search for HNL
decays into $\nu\mu^+\mu^-$ final states~\cite{ArgoNeuT:2021clc}. They derive limits on the mixing matrix element $\lvert U_{\tau 4}\rvert^2$.

The MicroBooNE collaboration has also published a search for HPS decaying to $e^+e^-$ pairs assuming the HPS originate from kaons decaying at rest after having been produced at the NuMI absorber~\cite{UbooneHPSPaper}. 
A data set corresponding to $1.93 \times 10^{20}$~POT is used to set limits on $\theta^2$ in the range $10^{-6}$--$10^{-7}$ at the $95\%$ CL for the mass range directly below the range considered in this paper ($0 < \mhps< 211$~MeV). Other direct experimental searches for HPS have been published in Refs.~\cite{NA62:2021bji,NA62:2021zjw,BNL-E949:2009dza_HPS,LHCb:2015nkv,LHCb:2016awg}, and reinterpretations 
of experimental data as HPS limits have been derived in Refs.~\cite{Winkler:2018qyg,Clarke:2013aya,PS191_reint,LSNDReint}.

\section{Heavy Neutral Leptons}
\label{sec:properties}

The HNL is introduced through an extension of the PMNS matrix by adding a single heavy mass eigenstate that mixes very weakly with the three active neutrino states. This minimal extension
adds four parameters to the model comprising the HNL mass $\mhnl$ and the elements of the extended PMNS matrix, $\lvert U_{\alpha 4}\rvert^2$ ($\alpha$ = $e$, $\mu$, $\tau$). The flavor eigenstates are 
\begin{equation}
    \nu_{\alpha} = \sum_i U_{\alpha i} \nu_i + U_{\alpha 4} N,
\end{equation}
with a heavy neutral lepton state $N$.
The HNL production and decay rates are suppressed by the relevant $|U_{\alpha4}|^2$ element through mixing-mediated interactions with SM gauge bosons. 

\begin{figure}[htbp]
\includegraphics[width=0.45\textwidth]{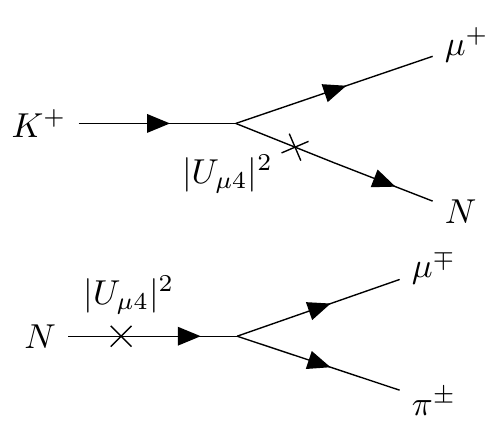}
\caption{Production of a Majorana HNL state $N$ via mixing in a $K^+$ meson decay and its subsequent decay into a $\mu^{\mp} \pi^{\pm}$ pair. A Dirac HNL will only decay into $\mu^-\pi^+$ pairs.}
\label{fig:HNL_decay1}
\end{figure}

HNLs would be produced in the decays of charged kaons and pions originating from the proton interactions on the targets of the BNB or NuMI neutrino beams. If the HNL lifetime is sufficiently long to allow the HNL to reach the MicroBooNE detector, they can decay into Standard Model (SM) particles within the argon volume. 

We consider the production channel $K^{+}\rightarrow \mu^{+} N$
with the decay $N \rightarrow \mu^{\mp} \pi^{\pm}$ as shown in Fig.~\ref{fig:HNL_decay1}.
The HNL production rate and decay width into $\mu \pi$ are each proportional to $\lvert U_{\mu 4}\rvert^2$, and the total rate therefore to $\lvert U_{\mu 4}\rvert^4$,
assuming $\lvert U_{e 4}\rvert^2 = \lvert U_{\tau 4}\rvert^2 = 0$~\cite{Atre:2009rg}. We thus place limits exclusively on the \umu{} mixing matrix element.
The accessible HNL masses are constrained by the condition $m_K - m_{\mu}>m_{\rm HNL}>m_{\mu} + m_{\pi}$.

\begin{figure}[htbp]
\includegraphics[width=0.46\textwidth]{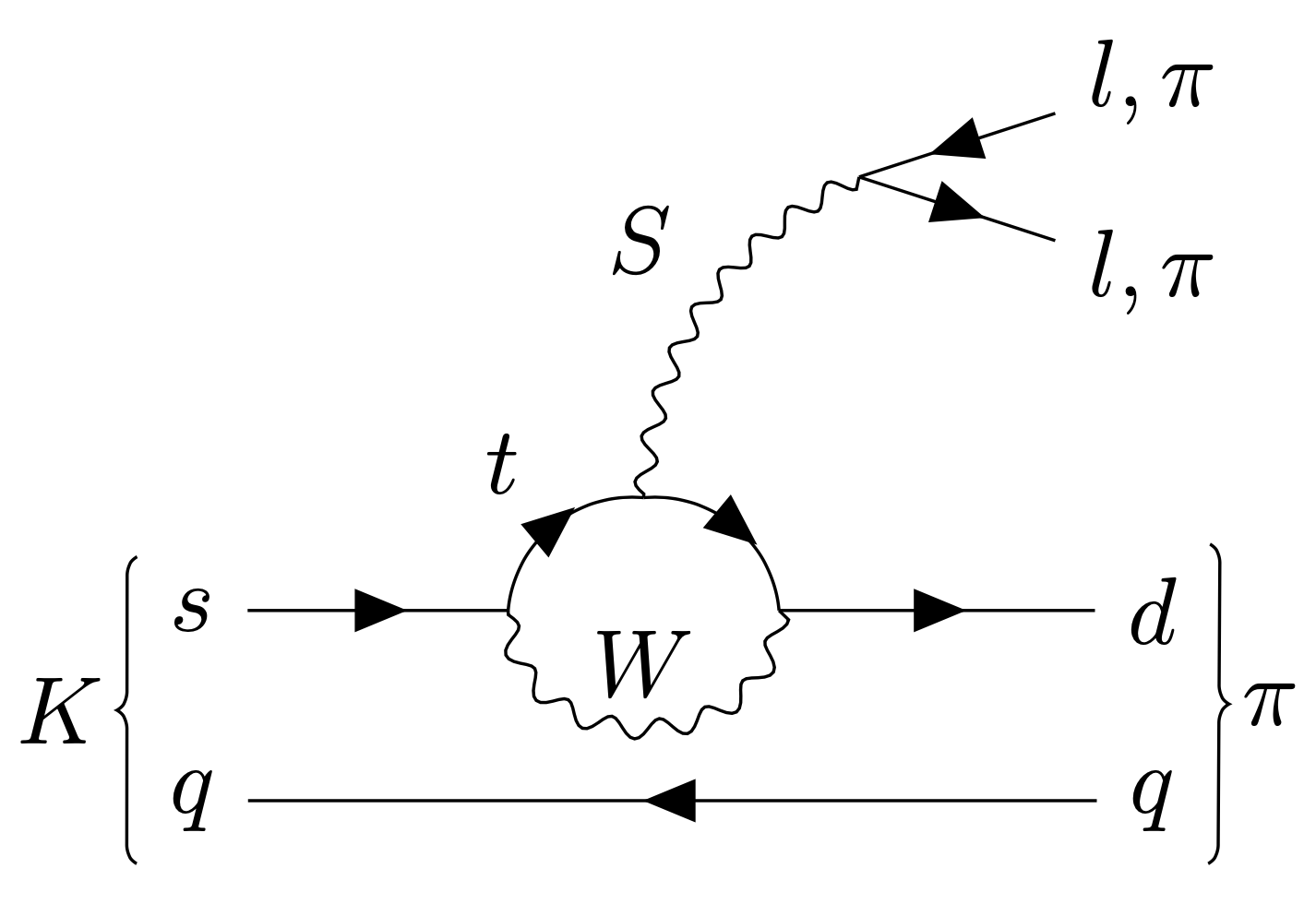}
\caption{Production of an HPS boson $S$ via mixing in a kaon meson decay and its subsequent decay. Only $K^{+}$ mesons contribute to this analysis, although this decay mode can also occur for neutral or $K^{-}$ mesons.} 
\label{fig:HPSprod}
\end{figure}

HNL states can include Dirac and Majorana mass terms, where
Majorana HNLs would decay in equal numbers into $\mu^+\pi^-$ and $\mu^-\pi^+$ final states. 
Dirac HNLs from $K^+$ decays could only decay to the charge combination $\mu^-\pi^+$ to conserve lepton number.

\section{Higgs Portal Scalars}
The Higgs portal model~\cite{patt:2006} is an extension to the SM, where an electrically-neutral singlet scalar boson mixes with the 
Higgs boson with a mixing angle $\theta$. 
Through this mixing, this HPS boson
acquires a coupling to SM fermions via their Yukawa couplings to 
the Higgs boson, which is proportional to $\sin\theta$. 
The phenomenology of the Higgs portal model at the SBN program, including the equations describing production and decay of the scalar boson, are discussed in Ref.~\cite{hpsbn}. 
At the absorber the dominant production channel will be through the two-body decay $K^+\rightarrow \pi^+S$ (where the HPS is denoted by $S$).  
The dominant decay mechanism is a penguin diagram with a top quark contributing in the loop (Fig.~\ref{fig:HPSprod}). 

The partial width for decays to charged leptons with mass $m_{\ell}$ is proportional to $m_{\ell}^2$~\cite{hpsbn}. If we assume that there are no new dark sector particles with masses $<\mhps/2$,
the branching fraction into $\mu^+\mu^-$ pair is therefore close to $100\%$
for scalar masses in the range $m_{\mu^+\mu^-}<\mhps<m_{\pi^0\pi^0}$.
The decays into $\pi^0\pi^0$ pairs become accessible for $\mhps>269$~MeV and
the decays into $\pi^+\pi^-$ pairs at $\mhps>279.1$~MeV.
 
We do not consider decays to neutral pion pairs in this search, since the detector signature differs significantly from muon and charged pions. The $\pi^+ \pi^-$ decay signatures appear very similar to $\mu^+\mu^-$ decays.
However, the analysis is not sensitive to HPS decays in this decay channel, as
the HPS will decay before it reaches the detector due its short lifetime.
We set limits as a function of the mixing angle $\theta$ as  
both HPS production and decay rate are proportional to $\theta^2$. 

\section{MicroBooNE Detector}
\label{sec:detector}

The MicroBooNE detector~\cite{Acciarri:2016smi} is a liquid-argon time projection chamber (LArTPC)
 situated at near-ground level at Fermilab. It is located at an angle of $8^\circ$ relative to the NuMI beamline~\cite{ADAMSON2016279} and at a distance of $680$~m from the target. 
 The MicroBooNE LArTPC has an active mass of $85$~t of liquid argon, in a volume $2.6 \times 2.3 \times 10.4$~m$^3$ in the $x$, $y$, and $z$ coordinates of the MicroBooNE coordinate system~\footnote{The MicroBooNE detector is described by a right-handed coordinate system, where the $x$ axis points along the negative drift direction with the origin located at the anode plane, the $y$
 axis pointing vertically upward with the origin at the center of the detector, and the $z$ axis
 points along the direction of the BNB beam, with the origin at the upstream edge of the detector. The polar angle is defined
 with respect to the $z$ axis and the azimuthal angle $\phi$ with respect to the $y$ axis.}.

 Charged particles produced in neutrino interactions with argon or in decays of LLPs will ionize the argon atoms along their trajectories, producing ionization electrons and scintillation light. An electric field of $273$~V/cm causes the electrons to drift towards the anode plane, requiring $2.3$~ms to drift across the width of the detector. 

The anode plane is oriented perpendicular to the electric field and comprises three planes of sense wires with a spacing of $3$~mm between adjacent wires and the same spacing separating the wire planes.
 Ionization electrons induce a bipolar signal when they pass through the first two planes of wires, oriented at $\pm 60^\circ$ with respect to the vertical,
before being collected on the third plane with vertically oriented wires producing a unipolar signal.

The waveforms measured by the $8192$ wires are digitized in a $4.8$~ms readout window.
This is longer than the $2.3$~ms drift time to allow reconstructing out-of-time cosmic rays.
The signal processing on the raw Time Projection Chamber (TPC) waveforms includes noise filtering and deconvolution to convert wire signals into hit information~\cite{Acciarri:2017sde,Ionization_processing}.
 Subsequently, individual hits corresponding to a localized 
 energy deposit are extracted for each wire.
 The combination of timing information and energy deposit contained in each waveform is used to create 2D projective views of the event.

An array of 32 8-inch photomultiplier tubes (PMTs) collects the scintillation light produced by argon ionization. 
Light flashes are reconstructed from the waveforms of the 32 PMTs.
To record an event, the NuMI online trigger requires a scintillation light signal in time with
the accelerator beam spill window above a configurable threshold for the number of photoelectrons, which was $9.5$ for Run~1 and then  lowered to $5.75$ during Run~3.

A cosmic ray tagger (CRT) surrounding the cyrostat was installed about midway through MicroBooNE operations \cite{MicroBooNE:CRT}. It comprises four panels made up of interleaved plastic scintillator strips placed above, below and on the two sides parallel to the BNB beam direction. The CRT provides both fast timing and positional information of cosmic rays entering the TPC.

\section{NuMI beamline}
  \begin{figure*}[htbp]
    \includegraphics[width=0.95\textwidth]{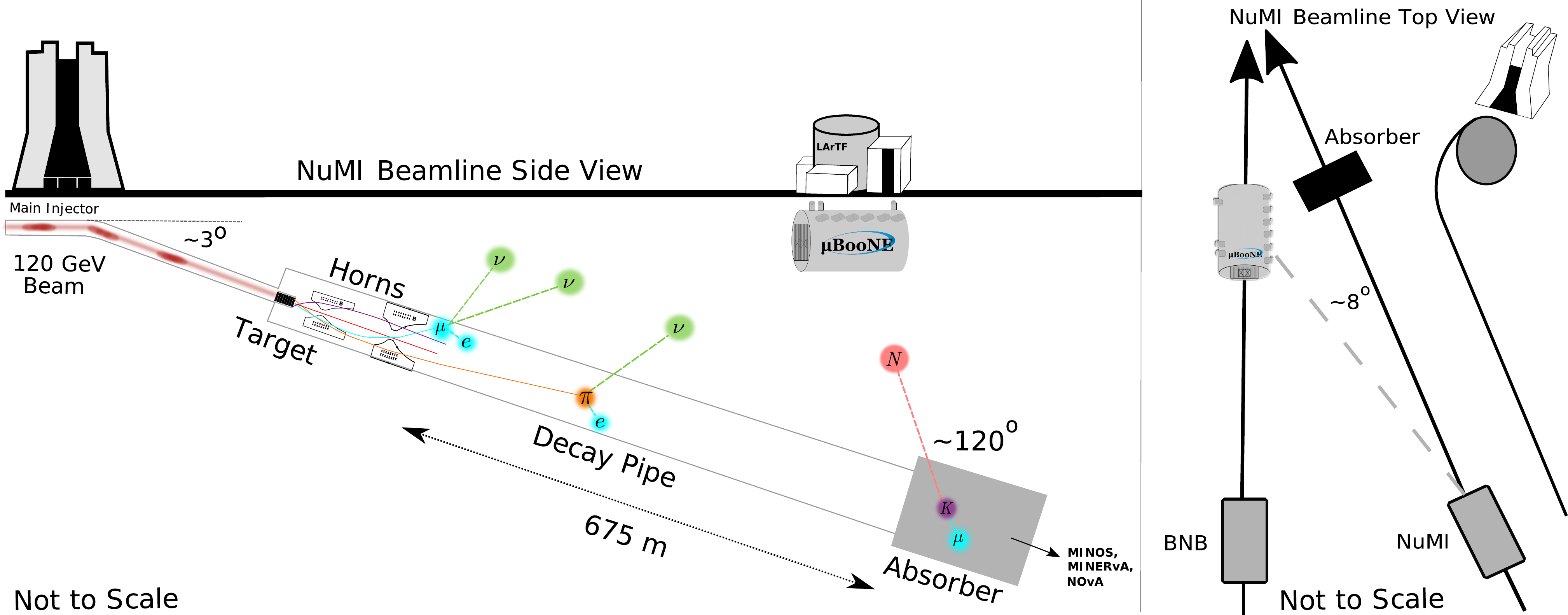}
    \caption{Illustration of the position of the MicroBooNE detector relative to the NuMI target and absorber.}
    \label{fig:NuMIdiagram}
\end{figure*}

 Protons with an energy of $120$~GeV from the Main Injector hit a graphite target, producing particles in the NuMI beamline. The position of the MicroBooNE detector relative to the components of the NuMI beamline is shown in Fig.~\ref{fig:NuMIdiagram}. A system of electromagnetic horns focuses the charged particles either towards or away from the beam axis, depending on the horn polarity. In forward horn current (FHC) mode, a positive ($+200$~kA) current is applied to the horns, which focuses positively charged particles in the beam direction. In reverse horn current (RHC) mode, a negative current ($-200$~kA) is applied to focus negatively charged particles. In this paper, we use NuMI data collected in both modes.

The focused particles then travel down a $675$~m long decay pipe filled with helium where they decay to neutrinos or anti-neutrinos. 
The proton beam structure determines the intensity and timing structure of the neutrino beam. The neutrino beam has six batches which together form a 
spill. Each spill is $9.6$~$\mu$s long.

Immediately downstream of the decay pipe, 
at a distance of $\approx 104$~m from the MicroBooNE detector, is an absorber ($5.5$~m wide, $5.6$~m tall, and $8.5$~m deep) made of an aluminium core surrounded by steel and then concrete, designed to absorb the remaining hadrons. 
In the MicroBooNE coordinate system, the direction from the absorber corresponds to $\theta= 2.20$ and $\phi= 1.15$ (in radians). 
Approximately $13\%$ of the beam protons pass through the target without interacting, travelling along the decay pipe before colliding with the absorber at a distance of $725$~m downstream from the target~\cite{ADAMSON2016279}.
These collisions produce a large number of $K^+$ mesons which then decay at rest, while most of the $K^-$ mesons
are absorbed. The LLPs studied in this paper would be produced in this absorber from $K^+$ decays. We assume equal rates of $K^+$ production for the two horn polarities~\cite{MiniBooNEKDAR}.

\section{Flux Generation} \label{sec:flux}

The flux of neutrinos in the NuMI beam is simulated in several steps as described in Ref.~\cite{MicroBooNENuMInuexsec1}. 
The NuMI beamline simulation uses the \texttt{g4Numi} code~\cite{Aliaga:2016oaz}, which is based on a \texttt{GEANT4} description of the geometry. The \texttt{PPFX} software package provides a neutrino flux prediction and uncertainties on the flux~\cite{PPFXPpaper}. 
We need to estimate the number of kaons decaying at rest in the absorber, in order to simulate the LLP signal with the \texttt{g4Numi} beamline simulation.

The NuMI flux from the absorber leads to a negligible rate of neutrino interactions contributing to, e.g., MicroBooNE cross-section results~\cite{MicroBooNENuMInuexsec1,MicroBooNENuMInuexsec2}.
The modeling of the flux of neutrinos from kaons decaying at rest in the absorber is however relevant for the LLP signal simulation. We therefore improve on this flux simulation by incorporating previous work from the MiniBooNE collaboration. 

The MiniBooNE detector is located downstream from the MicroBooNE detector in the NuMI beam. The MiniBooNE collaboration measured the $\nu_{\mu}$ flux from kaons decaying at rest at the NuMI absorber~\cite{MiniBooNEKDAR} using several methods
and compared it to several predictions. 
They adopted the \texttt{GEANT4} prediction of $0.085$ $\nu_{\mu}$ produced per POT as their central value, with a $30\%$ uncertainty taken from the range between simulations.
The flux predicted by \texttt{g4Numi} is almost an order of magnitude smaller than the MiniBooNE central value and is not consistent with the $30\%$ uncertainty. 
The \texttt{g4Numi} flux would yield a neutrino cross section measurement with the MiniBooNE detector that is inconsistent with expectations.
We therefore use the MiniBooNE central flux value for the
signal simulation, consistent with the procedure in Ref.~\cite{UbooneHPSPaper}.

\section{Signal Kinematics} \label{sec:signal}

We generate the LLP signal using the flux of charged kaons that
produce neutrinos, decaying them instead into an HNL or an HPS through the processes $K\rightarrow \mu N$ or $K\rightarrow \pi S$. The
two-body decay is isotropic in the kaon's rest frame, and the
energy $E_{\textrm{LLP}}$ of the LLP is given by
\begin{equation}
  E_{\textrm{LLP}}=\frac{m_{K}^2+\mllpsq-m_{\mu,\pi}^2}{2m_{K}},
  \label{eq:twobodyequation}
\end{equation}
where $\mllp$ is the LLP mass, $m_{K}$ the kaon mass, and $m_{\mu,\pi}$ the mass of the daughter particle, which is either a muon or a pion. 

The HPS decays into the $\mu^+\mu^-$ final state are simulated to reproduce the branching fraction as a function of HPS mass, with an isotropic decay
distribution in the HPS rest frame.
We simulate both charge conjugations of HNL decays, i.e.,~$\mu^{-}\pi^{+}$ and $\mu^{+}\pi^{-}$ final states, again with isotropic angular distributions. The angular distributions are then re-weighted for Dirac and Majorana HNL decays~\cite{Balantekin:2018ukw}.

 We assume a uniform time distribution of the proton beam interacting in the target, with a NuMI beam window of $9.6$~$\mu$s.
The two daughter particles are boosted to the lab frame using the LLP momentum vector.
The time of the LLP decay is calculated from the kaon decay time and the time of flight of the LLP to the decay point. All times of flight are taken into account 
in the simulation.

The LLP is only retained if its momentum vector intersects the detector volume.
A decay vertex within the detector volume is then chosen along the trajectory of the LLP. 
The exponential decay of the HPS flux is accounted for when selecting a decay vertex. The HPS lifetime is proportional to $\theta^{-2}$~\cite{hpsbn}.
The decay length of an HPS with $m_{\rm HPS}>2m_\mu$ and a mixing angle in the region of
interest of $10^{-7}<\theta< 10^{-9}$ is similar to the distance between the absorber and the MicroBooNE detector. Therefore some HPS will decay before reaching the detector, reducing the flux in the MicroBooNE detector.

For large values of $\theta^2$, only a small fraction of the HPS reach the detector before decaying, which restricts the upper reach of exclusion contours as a function of $\theta^2$.
To derive these contours, we first define an effective angle $\theta_{\textrm{eff}}$ which neglects the impact of decays before the 
detector. 
The event rate in the MicroBooNE detector is then $\propto\theta_{\textrm{eff}}^4$. In the final extraction of the limits, we correct for the early decays to obtain the limits as a function of the physical mixing angle $\theta^2$.

The exponential decay of the HNL flux is negligible for the mixing angles considered here
as the HNL lifetime is much longer than the time needed to reach the MicroBooNE detector. The number of HNL decaying before reaching the detector is therefore neglected, and the final event rate is proportional to $\lvert U_{\mu 4}\rvert^4$.

\section{Simulation and Reconstruction} \label{sec:mc}

Simulation and reconstruction are performed within the \texttt{LArSoft}~\cite{Snider:2017wjd} framework.
 The neutrino event generator \texttt{GENIE}~\cite{GENIEcite} simulates the neutrino interactions on argon inside the cyrostat as well as interactions with the surrounding material. The \texttt{GENIE} configuration used in the simulation is found in Ref.~\cite{MicroBooNEGenieTune}, which describes a tuning of phenomenological parameters related to charged-current quasi-elastic
scattering and scattering on a pair of correlated nucleons, based on a fit to external data. 

To obtain the response of the detector to ionization charge and scintillation light, 
the propagation through the detector material
of secondary particles produced in the LLP decays or in the neutrino interactions is simulated by \texttt{GEANT4}~\cite{Agostinelli:2002hh}.

Cosmic rays crossing the detector in the same readout window are taken from data recorded outside of the beam window and then overlaid on 
the MC simulation. This overlay also addresses the need for simulating detector noise, as the simulated waveforms are combined with waveforms from data recorded during beam-off times.

The simulated samples and the data are reconstructed using the same algorithms.  We use the \texttt{Pandora} pattern recognition framework~\cite{Marshall:2015rfa} to combine hits, first clustered independently on each anode plane and then combined across anode planes to build particles reconstructed in 3D. The particles are arranged in a parent-daughter hierarchy based on the topology of the event and classified as track-like (muons, charged pions, and protons) or shower-like (electrons and photons). 

Optical hits are constructed from PMT waveforms. Time-coincident optical hits from different PMTs are combined to form “flashes” that are attributed to a single interaction in the detector. The time of the flash, the associated location, and the amount of light are determined for each flash. 

\section{Data Sets} \label{sec:data}

In total, we recorded data corresponding to $2.23\times 10 ^{21}$~POT with the MicroBooNE detector exposed to the NuMI beam. 
For this paper, we analyse a subset of the data corresponding to $7.01 \times 10^{20}$~POT, which were taken in two different operating modes, forward horn current (FHC) and reverse horn current (RHC).
The FHC data set was recorded during Run 1 in 2015--2016 and the RHC data set during Run 3 in 2017--2018. 
The CRT was fully installed for the second period, where it is used in the analysis.
We thus analyze the two data sets separately to account for differences in neutrino flux, detector configuration, and CRT coverage.

The ``beam-on" samples are defined by triggers that are coincident with the neutrino beam; they are used to 
search for the LLP signal. In addition, we use three samples for each each data period (FHC and RHC) that
are designed to describe the background (``beam-off", ``in-cryostat neutrino", and ``out-of-cryostat neutrino"). 

The majority of beam-on events do not contain a neutrino interaction in MicroBooNE and are triggered by a cosmic ray.
This source of background is modeled using a sample of events collected under identical trigger conditions but at times where no neutrino beam is present. These samples are referred to as ``beam-off". The beam-off sample is scaled so that its normalisation corresponds to the number of beam spills of the beam-on sample.
An additional scaling factor of $0.98$ is applied to the beam-off sample.
This factor takes into account that about $2\%$ of all NuMI spills 
contain a neutrino interaction in the detector, and the remaining spills contain only cosmic rays~\cite{krish}.

The neutrino-induced background from the NuMI beam is modeled using two samples for each run period. The ``in-cryostat $\nu$" sample contains interactions of neutrinos with the argon inside the cryostat, and the ``out-of-cryostat $\nu$" sample describes interactions with the detector structure and surrounding material. Both samples are generated with the MC simulation and normalized to the number of POT. The number of POT and the number of events passing the online-trigger before scaling factors are applied are summarized in Table~\ref{tab:samples}.
\begin{table}[ht!]
\centering
 \setlength{\tabcolsep}{7.0pt} 
\renewcommand{\arraystretch}{1.15}
\caption{Number of events before applying scaling factors for the data and background samples, separated by run periods.
The corresponding number of POT are given for the beam-on data and for the simulated background samples (in-cryostat $\nu$ and out-of-cryostat $\nu$).
}
\label{tab:samples}
\begin{tabular}{cccccc} 
    \hline\hline
& \multicolumn{2}{c}{Run 1 (FHC)}  &
\multicolumn{2}{c}{Run 3 (RHC)}  \\
    & POT &   events & POT &   events \\
         &     $\times 10^{20}$  & $\times 10^5$ & $\times 10^{20}$  & $\times 10^5$\\
    \hline
\multicolumn{2}{l}{data sample:}  & & & \\
beam-on  & $2.00$ & $\phantom{0}6.11$ & $5.01$ &   $11.04$  \\
\multicolumn{2}{l}{background samples:} & & & \\
 beam-off &  N.A. & $\phantom{0}9.12$ & N.A. &   $32.37$\\
in-cryostat $\nu$ & $23.3$ &   $\phantom{0}9.11$ & $19.8$ &   $\phantom{0}7.46$ \\
 out-of-cryostat $\nu$  &  $16.7$  &  $\phantom{0}5.69$
& $10.3$  &  $\phantom{0}3.86$
\\
\hline\hline
\end{tabular}
\end{table}

\begin{figure}[ht!]
  \centering
  \includegraphics[width=0.46\textwidth]{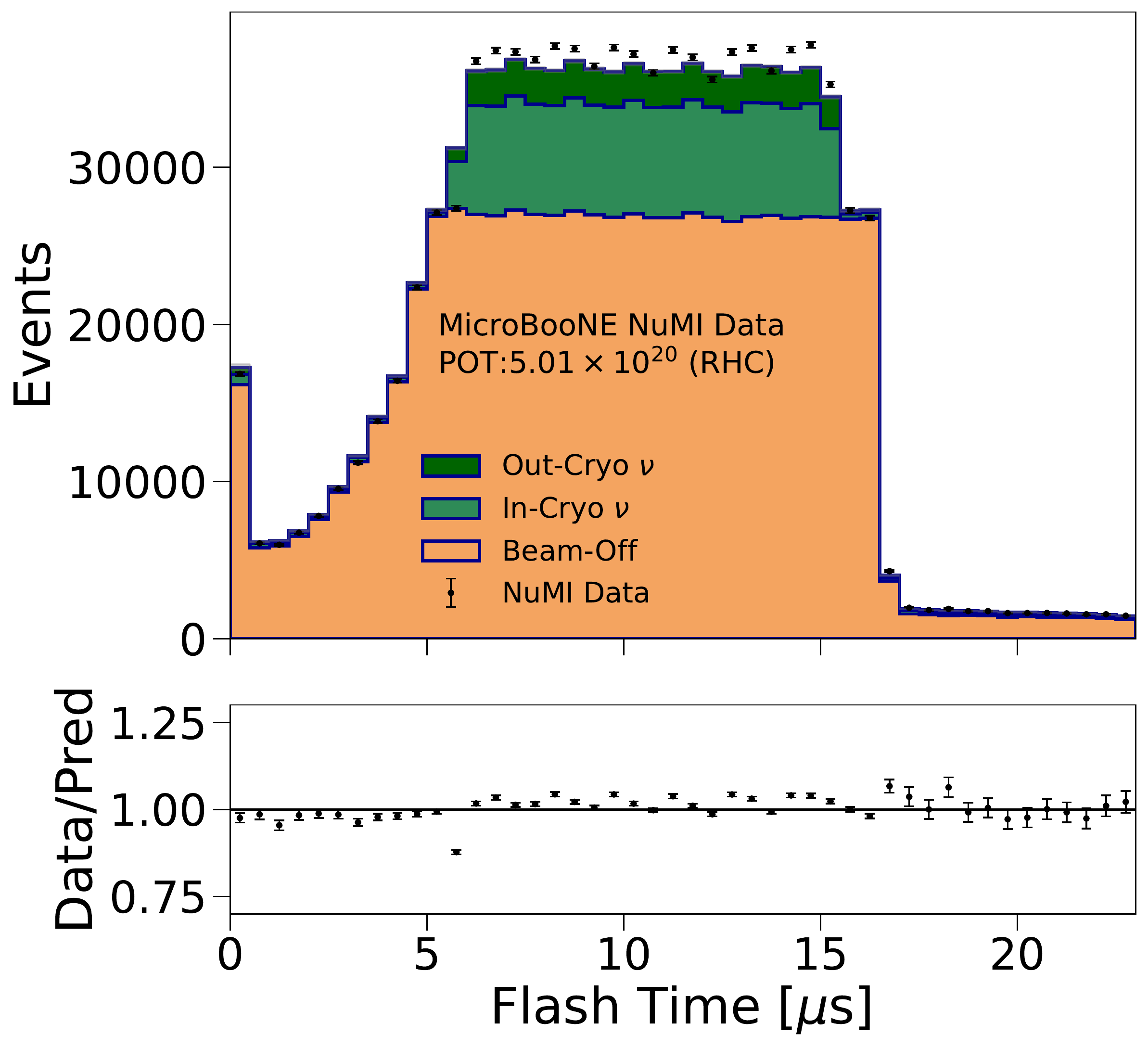}
  \caption{Distributions of flash times for all Run 3 data events. The beam-on data are compared to the sum of the beam-off, in-cryostat $\nu$, and out-of-cryostat $\nu$ samples. The single bin at the start of the neutrino spill at $5.64$~$\mu$s, where the expectation significantly exceeds the data, is caused by an effect in the timing structure of the beam that is not included in the simulation. The overall normalisation of the data and prediction agree within $\approx 2\%$.}
   \label{fig:Run3Flashtime}
\end{figure}
After matching the sample size to the number of POT of the beam-on data, the out-of-cryostat $\nu$ samples are scaled by an additional factor 
%of $0.75$ for the FHC sample and $0.35$ for the RHC sample 
to ensure that the sum of the 
beam-off, in-cryostat and out-of-cryostat $\nu$ samples matches the data normalization within the NuMI timing window of $5.64$--$15.44$~$\mu$s, as shown in Fig.~\ref{fig:Run3Flashtime}. 
There is a small residual difference of $\approx 2\%$ in the normalization of data relative to the sum of the background samples at this stage, as the scaling factors are derived with only a subset of the data. 

Simulated data sets using the procedure described in Sec.~\ref{sec:mc} are used to evaluate the reconstruction and selection efficiency for the signal LLP decays. 
In total, we generate twelve samples in the mass range $246\le m_{\rm HNL} \le 385$~MeV 
and eight samples in the mass range $212\le m_{\rm HPS} \le 279$~MeV.
The spacing of the mass points takes into account the resolution. Additional points were generated at the edges of phase space, 
where limits could change more rapidly.

\begin{figure}[htbp]
\includegraphics[width=0.49\textwidth]{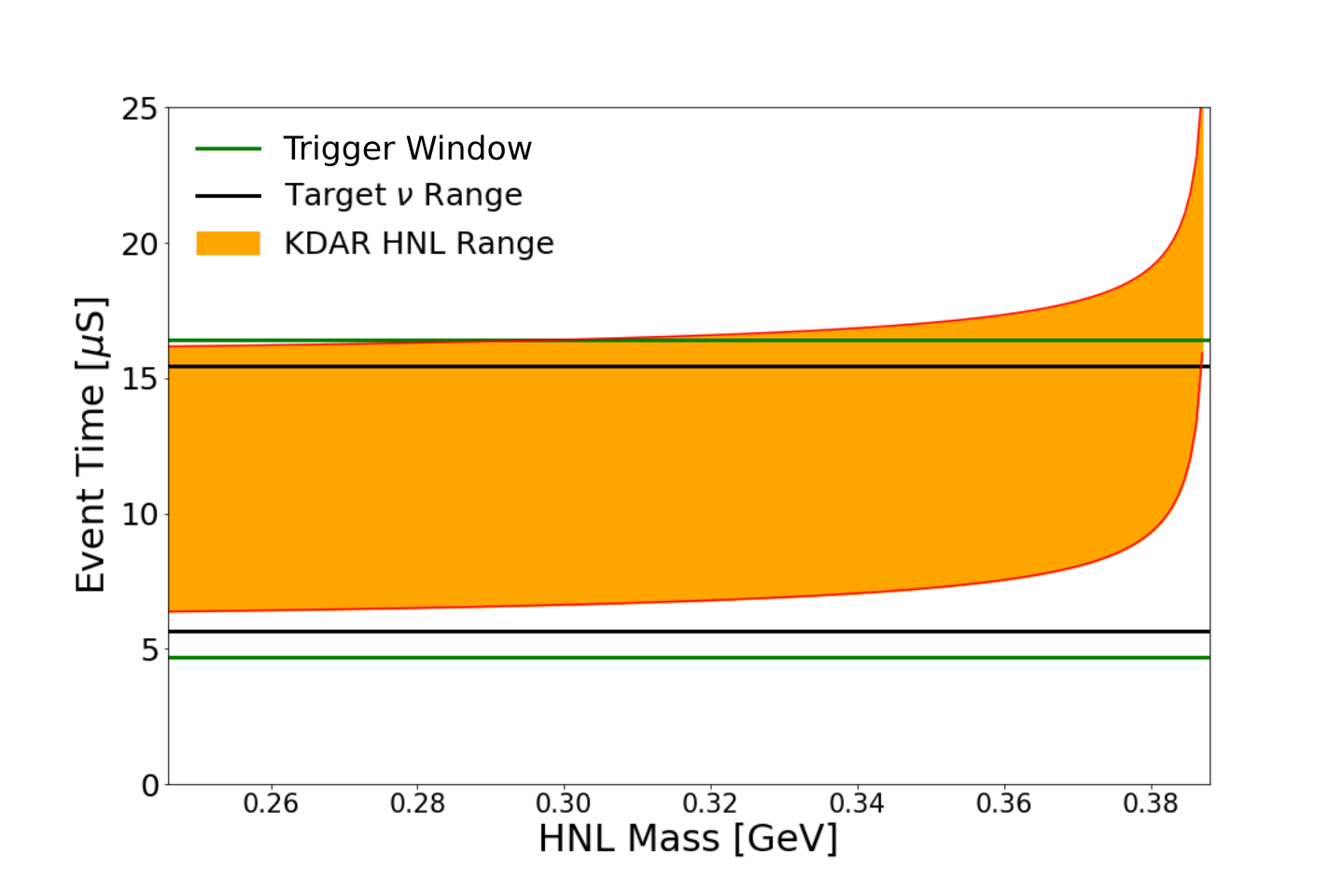}
\caption{Expected time of arrival of HNLs as a function of $m_{\rm HNL}$. The horizontal lines show the timing range
consistent with the expected time of arrival of neutrinos from the NuMI target. 
The trigger window extends by about $1$~$\mu$s on both sides compared to
the timing range for neutrino interactions.}
\label{fig:timing}
\end{figure}

The expected time of arrival of HNLs as a function of $m_{\rm HNL}$ is shown in Fig.~\ref{fig:timing}.
The MicroBooNE NuMI trigger requires a flash in a timing window of $4.69$--$16.41$~$\mu$s that covers the arrival times of the NuMI neutrino beam at the detector.
More than $95\%$ of HNLs arrive within the trigger window at low $\mhnl$.
Since the time of flight increases with $\mhnl$, the trigger efficiency decreases to 
$\approx 55\%$ for $\mhnl=385$~MeV and then quickly goes to zero. As we study a lower mass range for the HPS, this effect is not relevant for the HPS search. 

\section{Candidate Vertex Identification} 

The Pandora reconstruction algorithm groups objects
into “slices” after removing cosmic-ray related hits.
The slice
identification (SliceID) process uses a combination of charge and light information to identify whether a slice corresponds to a neutrino interaction.
Slices containing through-going, out-of-time, or stopping muons are removed. The reconstructed charge in the slice is required to be consistent with the location and intensity of the flash that has triggered the event.
If this selection leaves $\ge 2$ remaining slices in the 
event, a Support Vector Machine (SVM) is applied to calculate a ``topological score" to select the most “neutrino-like” of the slices. This slice is then also used for the LLP search.

The LLP topology is characterized by a single vertex with two decay particles. 
The \texttt{Pandora} reconstruction frequently misplaces the location of the LLP decay vertex as the signal topology differs from the standard neutrino interactions (see Figs.~\ref{fig:HNL_decay} and \ref{fig:sigeventdisplays}).
The vertex locations are therefore re-calculated for this analysis.
All pairs of fitted tracks whose start or end points lie within a 3D-distance of $5$~cm of each other are combined to form signal candidates with a common vertex.
If the end point of a track is placed at one of the new candidate vertices, the track direction is reversed. The vertex coordinates are calculated as the mean of the start locations of the two tracks. At this
stage, there can be multiple vertices in an event.
Approximately $(60-70)\%$ of signal events contain at least one vertex and $(40-50)\%$ contain exactly one vertex for the generated signal samples (Table~\ref{tab:TabVers}). 

\section{Kinematic and Topological Variables}
\begin{table*}[htbp!]
\caption{Number of events with at least one candidate vertex and the total number of vertices for events that pass the candidate selection, normalized to the number of POT of the beam-on data sample. 
The number of expected HNL candidates is scaled to $|U_{\mu4}|^2=10^{-8}$, and the number of HPS candidates to $\theta_{\textrm{eff}}^2=10^{-9}$. The percentages are calculated with respect to the number of events that pass the \texttt{Pandora} SliceID procedure.} 
\setlength{\tabcolsep}{8pt} 
\renewcommand{\arraystretch}{1.10}
\begin{tabular}{ *{5}{c} }
    \hline\hline
Sample     & \multicolumn{2}{c}{Run 1 (FHC)}
            & \multicolumn{2}{c}{Run 3 (RHC)}
              \\

                     &   \multicolumn{1}{c}{Events}  &  Vertices  &  
                      \multicolumn{1}{c}{Events}     & Vertices \\
    \hline
beam-off  &  $24033$ $(38.8\%)$   &  $37313$   &   $50068$ $(38.6 \%)$  &   $\phantom{0}77776$   \\
in-cryostat $\nu$      &   $17736$ $(58.3\%)$ &  $43226$  &   $45582$ $(59.8\%)$  &   $119443$ \\
out-of-cryostat $\nu$   &   $\phantom{0}3620$ $(30.1\%)$  & $\phantom{0}4952$ &   $\phantom{0}4557$ $(27.8\%)$  & $\phantom{00}6418$  \\
\hline
sum of predictions & $45390$ $(43.5\%)$ & $85490$ & $100207$ $(45.1\%)$ & $203636$  \\ 
beam-on (Data)                 &  $45638$ $(44.1\%)$   &  $86834$  &   $\phantom{0}98061$ $(44.1 \%)$  & $194260$  \\
\hline 
data/prediction      &   $1.01$  &  $1.00$  &   $0.98$ &   $0.95$ \\
\hline
signal & & & &\\
$\mhnl=304$~MeV &  $10.7$ $(64.7\%)$   &  $16.0$ &  $26.5$ $(64.6\%)$  &   $40.2$ \\
$\mhps=245$~MeV &  $\phantom{0}8.8$  $(64.4\%)$   &  $13.6$ &   $22.1$ $(66.7\%)$ & $34.8$ \\
    \hline\hline
\end{tabular}
\label{tab:TabVers}
\end{table*}
We define several kinematic and topological variables to discriminate between signal and background. These variables are either associated to the slice or to the signal candidate. The slice-related variables use information calculated
for the entire slice:
\begin{itemize}
 \item{\bf Multiplicity:} The total multiplicity of objects in the slice, $N_{\rm tot}$, and the multiplicities of the objects classified as either tracks or showers ($N_{\rm tr}$ or $N_{\rm sh}$). 
 \item{\bf Containment:}
    Events are required to be contained inside the TPC's active volume by restricting the maximum and minimum coordinates of the start and end points of the objects within the slice to be
\begin{eqnarray}
\nonumber
    9 < x <253~\text{cm},\\ \nonumber
    -112 < y <112~\text{cm},\\
    14 < z <1020~\text{cm}.
\end{eqnarray}
We also define the maximum and minimum extent of the slice for each coordinate, denoted as $\max{(i)}$ and $\min{(i)}$, with $i=x,y,z$. They are defined as the largest or smallest value of the start and end points of tracks in the slice in each dimension.
 \item {\bf Slice energy:}
        The energy of all the objects of the slice, $E_{\textrm{sl}}$, is reconstructed from the charge read-out by the collection planes. We expect that
    \begin{equation}
    E_{\textrm{sl}}\approx E_{\rm LLP}-(m_{1}+m_{2}),
    \label{eq:Esl}
    \end{equation}
    where $E_{\rm LLP}$ is the energy of the LLP and $m_{1}$ and $m_{2}$ are the masses of the two decay particles.
    This assumes the decay particles are stable and their kinetic energy is fully measured in the TPC. The ionization energy contributed by secondary decays will increase $E_{\rm sl}$.

\item{\bf Topological score:}
A Support Vector Machine (SVM) is applied to calculate a topological score that selects the slice that is most consistent with containing a neutrino interaction.
\end{itemize}
The remaining variables are directly related to the signal candidate.
\begin{itemize}
 \item{\bf Proton likelihood:}
The log-likelihood particle identification score, $S_{\rm PID}$, is designed to utilize the calorimetric information to discriminate between protons and minimally ionizing particles (muons)~\cite{MicroBooNELLR}. A score of $S_{\rm PID}=-1$ indicates that a track is consistent with the proton hypothesis and $S_{\rm PID}=+1$ that a track is consistent with the muon hypothesis. 
 \item{\bf Track length:}
We calculate the lengths $\ell_t$ of the two tracks defining the candidate in the TPC.
 \item{\bf Candidate four-momentum:}
The momenta of the particles associated to the two tracks produced by the LLP decay are determined using the length of the tracks and the continuous-slowing-down-approximation~\cite{csdamu}. For the HNL decays the momenta and candidate mass are calculated assuming that the longer track is the muon and the shorter track is the pion. 
The momenta are summed to calculate the LLP candidate's four-momentum. We define $\beta$ as the angle between the momentum of the candidate and the vector connecting the centre of the absorber to the candidate vertex.
The angle $\alpha$ is the opening angle of the two tracks. 
\end{itemize}

\section{Event selection}
\label{selection}

The first stage of the final selection requires that the flash time coincides with the beam window, and only the flash with the largest number of associated photoelectrons is used.  
If there is a CRT hit within $1$~$\mu$s of the flash for the Run 3 (RHC) sample, the event is identified as a cosmic ray and rejected. The 3D distance between the Pandora vertex and the closest tracks reconstructed using the CRT is required to be $>20$~cm to remove events
where charge from cosmic rays not in time with the beam flash is reconstructed as part of a candidate.

We require $2\le N_{\rm tot}\leq 4$, $N_{\rm tr}\leq3$, and $N_{\rm sh}\leq2$. The event must be contained and the reconstructed energy $E_{\rm sl}<500$~MeV. To remove neutrino interactions producing a proton, candidates where at least one track has a score $S_{\rm PID}<-0.5$ are removed. Candidates are required to have an opening angle $\cos{\alpha}>-0.94$. This removes events where a single cosmic-ray is split into two tracks with a large opening angle. Finally, the longest track must be shorter than $50$~cm.

\begin{table}[htb]
\caption{Number of events with at least one candidate vertex that pass in each sample after the full event selection. All numbers are POT normalized. HNL candidates are scaled to $|U_{\mu4}|^2=10^{-8}$, HPS to $\theta_{\textrm{eff}}^2=10^{-9}$. The efficiencies are given with respect to the number of events that pass the \texttt{Pandora} SliceID procedure.}
\setlength{\tabcolsep}{6pt} 
\renewcommand{\arraystretch}{1.05}
\begin{tabular}{ccccc}
\hline\hline sample & \multicolumn{2}{c}{Run 1 (FHC)} & \multicolumn{2}{c}{Run 3 (RHC)} \\
& events & efficiency & events & efficiency \\
 \hline beam-off & 1552 & $2.5 \%$ & 1234 & $0.9 \%$ \\
in-cryostat $\nu$ & 1188 & $3.9 \%$ & 2132 & $2.7 \%$ \\
out-of-cryostat $\nu$ & $\phantom{0}208$ & $1.7 \%$ & $\phantom{0}129$ & $0.8 \%$ \\
\hline 
sum of predictions & 2948 & $2.8 \%$ & 3495 & $1.6 \%$ \\
beam-on (data)  & 2950 & $2.8 \%$ & 3410 & $1.5 \%$ \\
\hline data/prediction & $1.00$ &  & $0.99$ & \\
\hline signal & & & & \\
 $\mhnl=304$~MeV & $7.5$ & $45.2 \%$ & $17.6$ & $43.1 \%$ \\
 $\mhps=245$~MeV & $6.2$ & $45.9 \%$ & $15.1$ & $45.6 \%$ \\
\hline\hline
\end{tabular}
\label{tab:presel}
\end{table}

Table~\ref{tab:presel} shows the combined efficiency of the selection requirements and vertex reconstruction 
relative to the SliceID requirements for background and
signal. The background rejection is better for Run~3 compared to Run~1 as the CRT improves the rejection of cosmic rays.
The CRT also vetoes particles produced in neutrino interactions that either enter or exit the detector, as expected for out-of-cryostat and in-cryostat neutrino interactions. The effect of the CRT veto on the signal is small, as the decay
tracks are shorter. 
The signal efficiency relative to the SliceID requirements are in the range $(36$--$45)\%$, rising with LLP mass. 
The total efficiency for an LLP decaying in the detector to be triggered, to be reconstructed and selected by the SliceID, and to pass the final event selection is in the
range $(13$--$30)\%$, again increasing with LLP mass.

\begin{figure*}[htbp]
  \centering
  \includegraphics[width=0.325\textwidth,trim={0.75cm, 0, 0.75cm, 1cm}, clip]{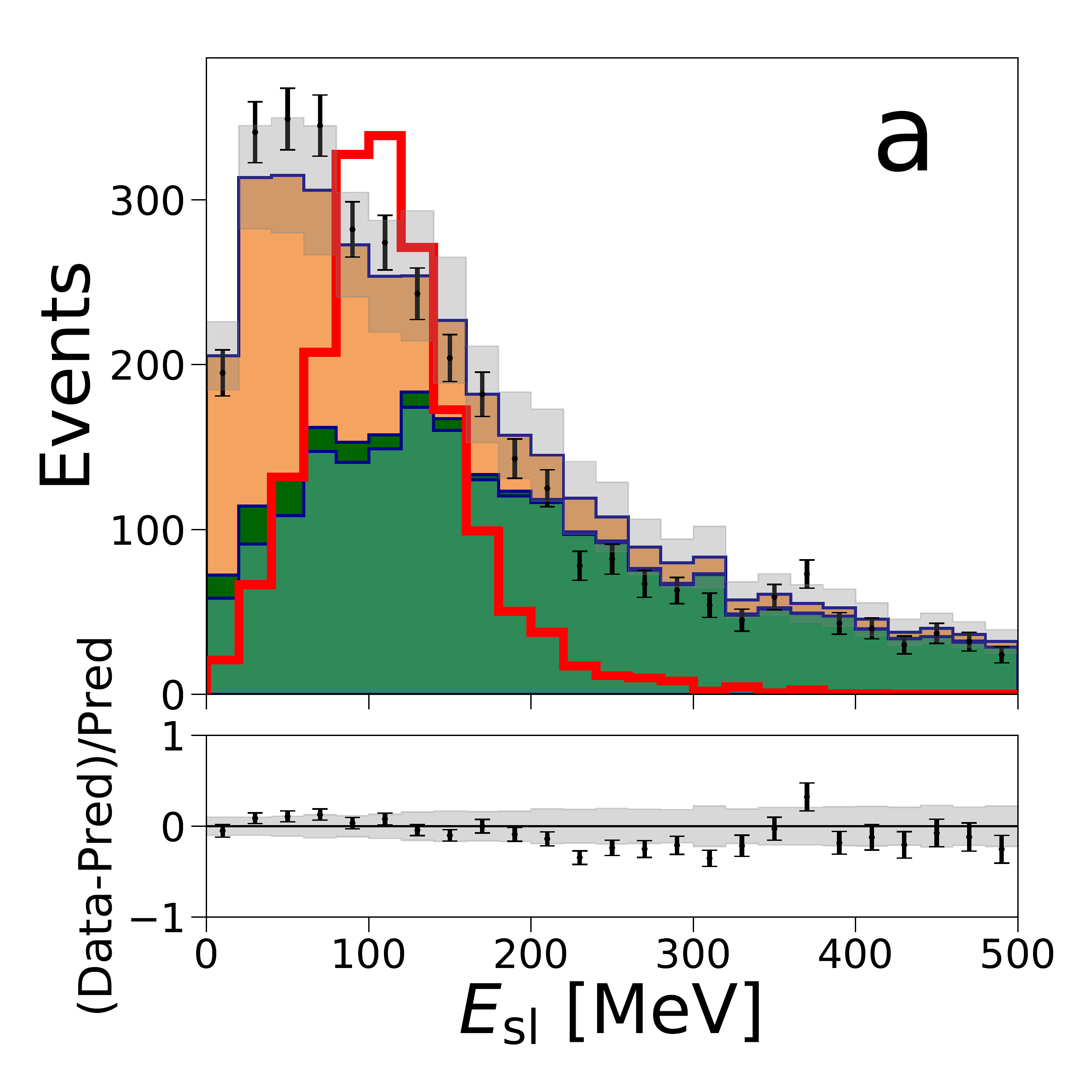}
  \includegraphics[width=0.325\textwidth,trim={0.75cm, 0, 0.75cm, 1cm}, clip]{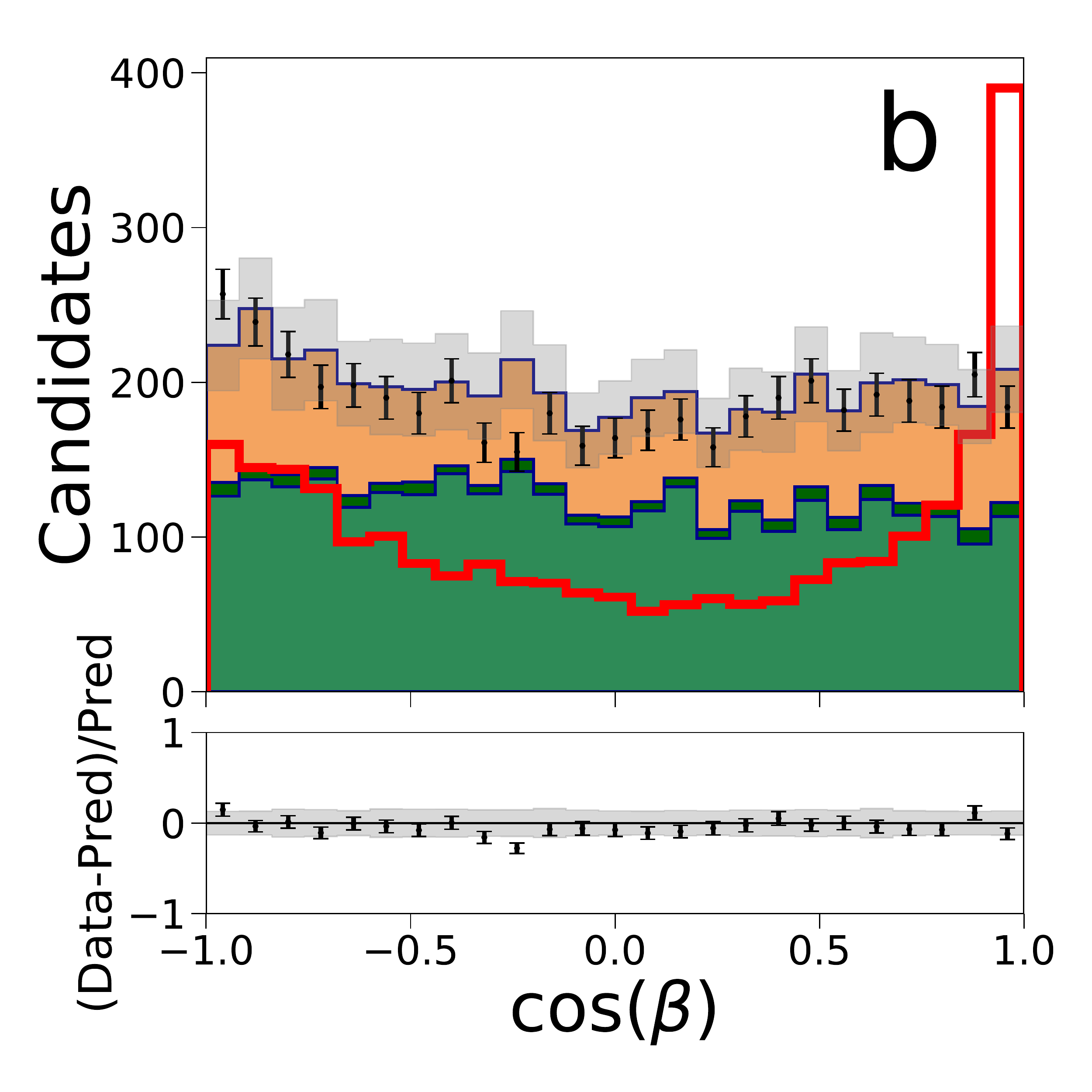}
    \includegraphics[width=0.325\textwidth,trim={0.75cm, 0, 0.75cm, 1cm}, clip]{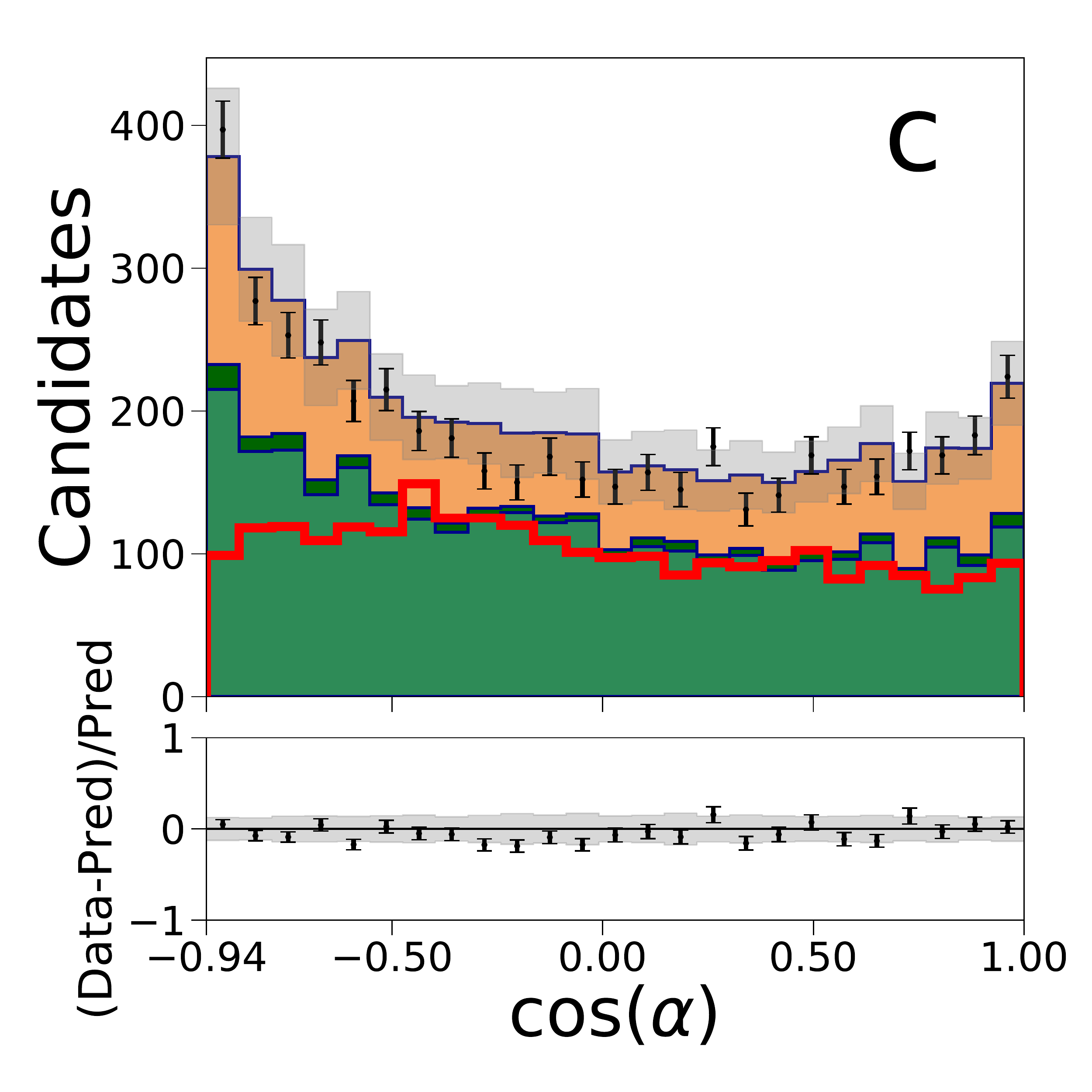}
  \includegraphics[width=0.325\textwidth,trim={0.75cm, 0, 0.75cm, 1cm}, clip]{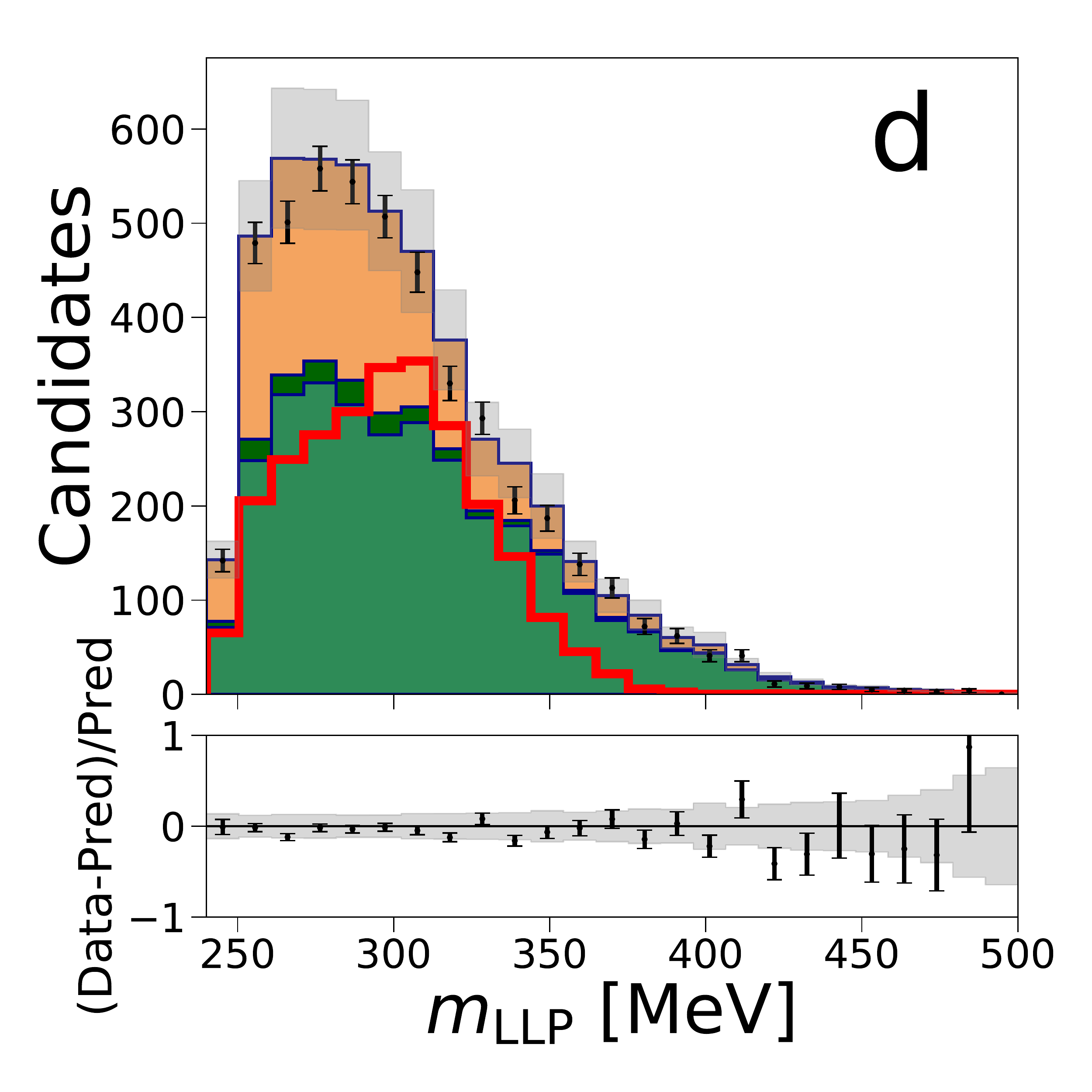}
      \includegraphics[width=0.325\textwidth,trim={0.75cm, 0, 0.75cm, 1cm}, clip]{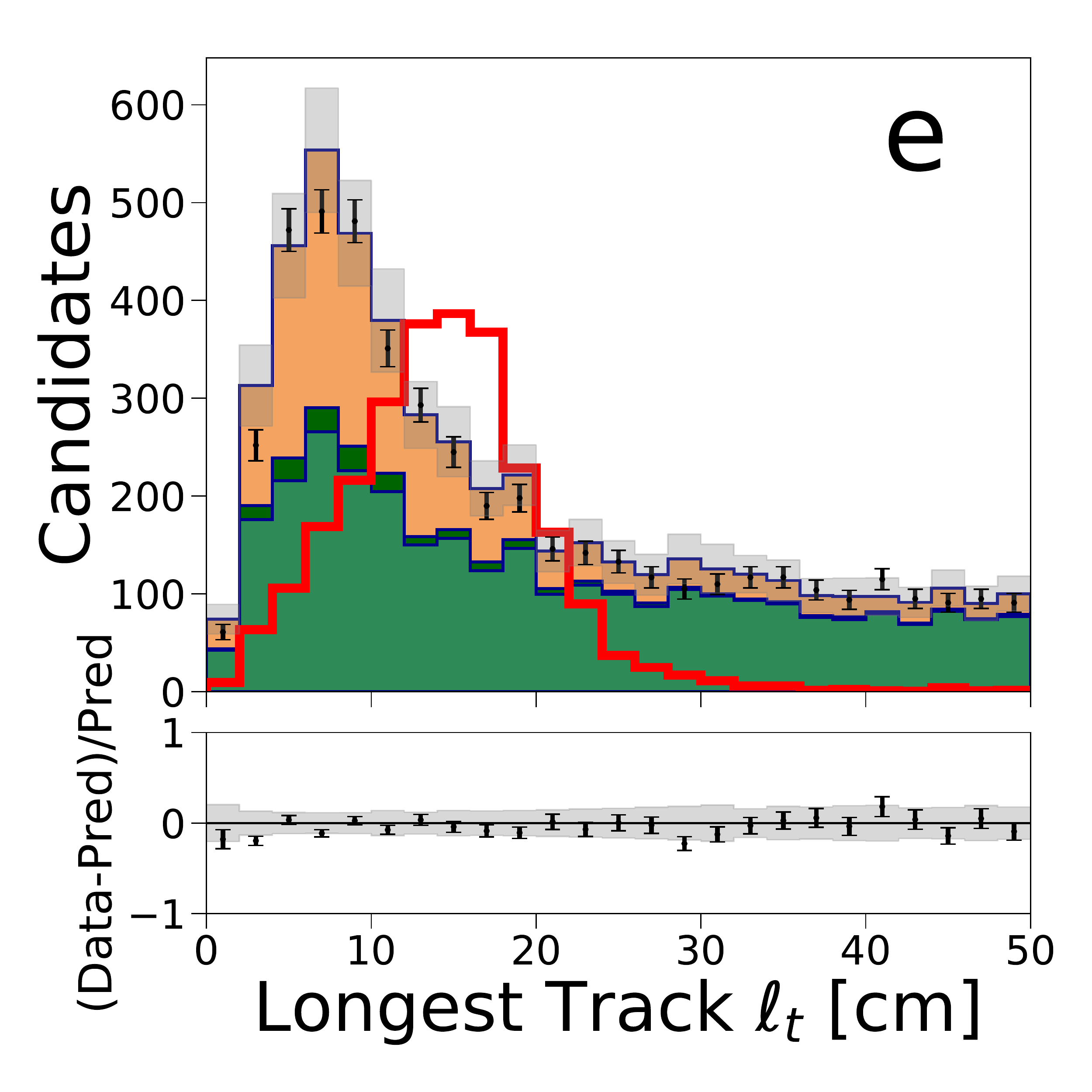}
  \includegraphics[width=0.325\textwidth,trim={0.75cm, 0, 0.75cm, 1cm}, clip]{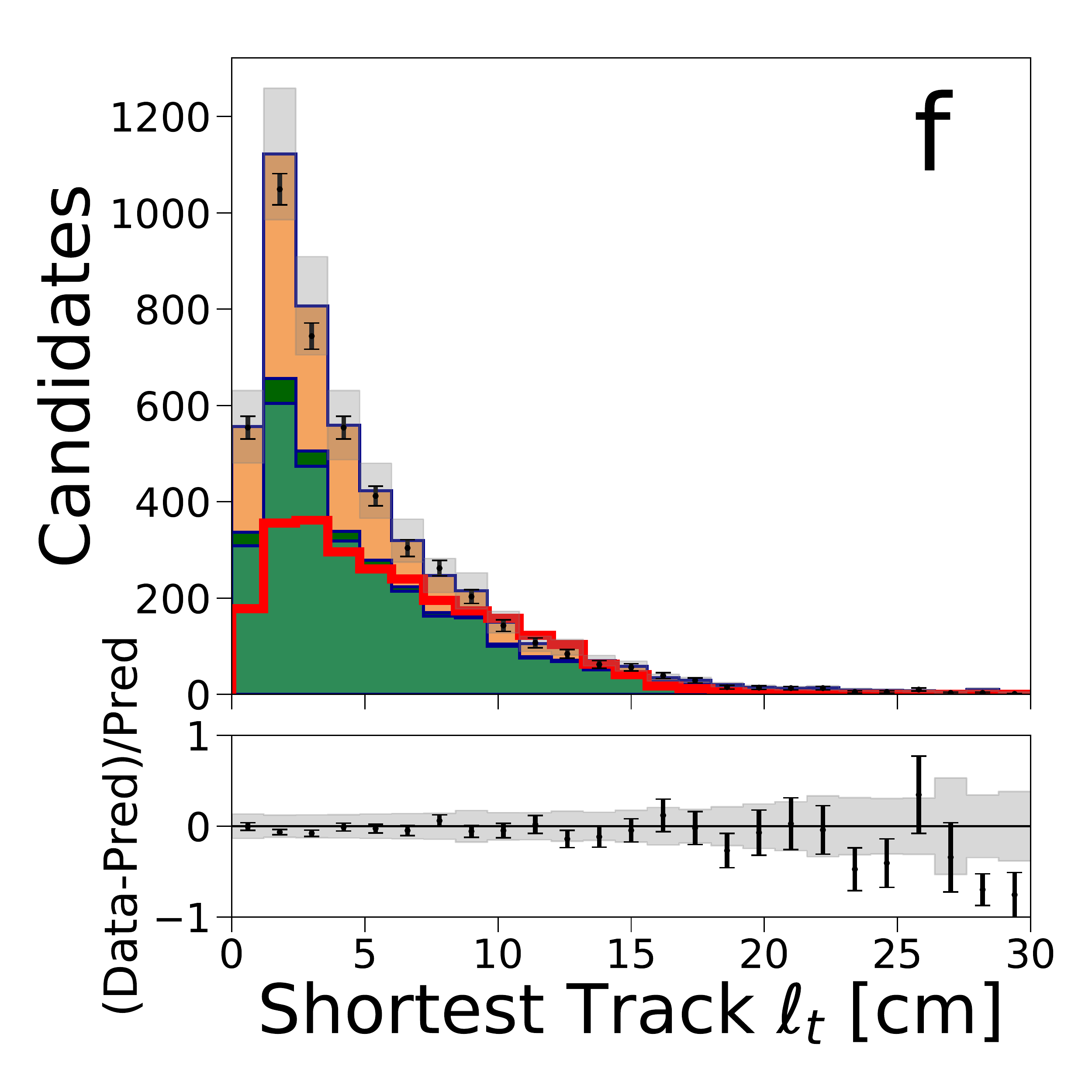}
    \includegraphics[width=0.325\textwidth,trim={0.75cm, 0, 0.75cm, 1cm}, clip]{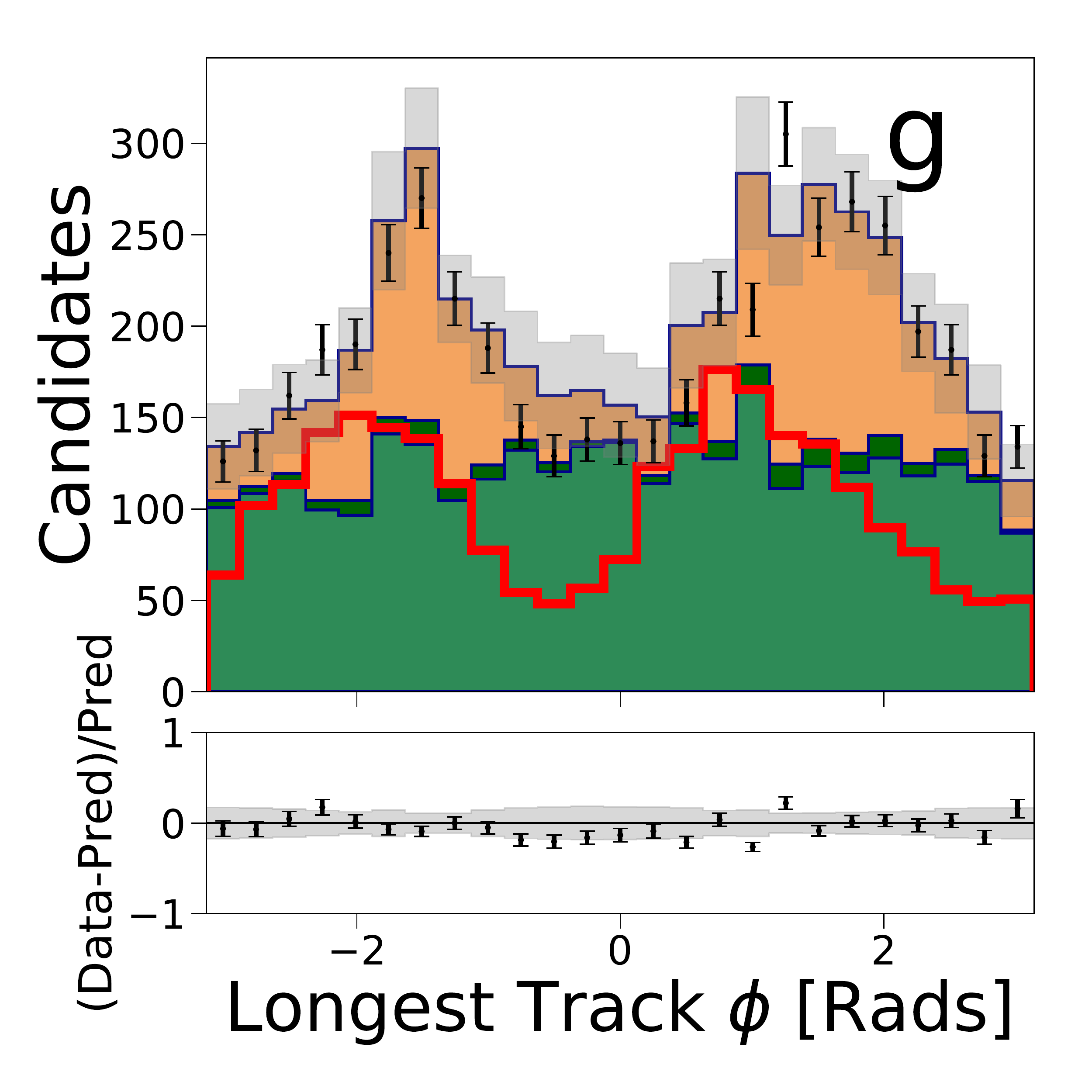}
  \includegraphics[width=0.325\textwidth,trim={0.75cm, 0, 0.75cm, 1cm}, clip]{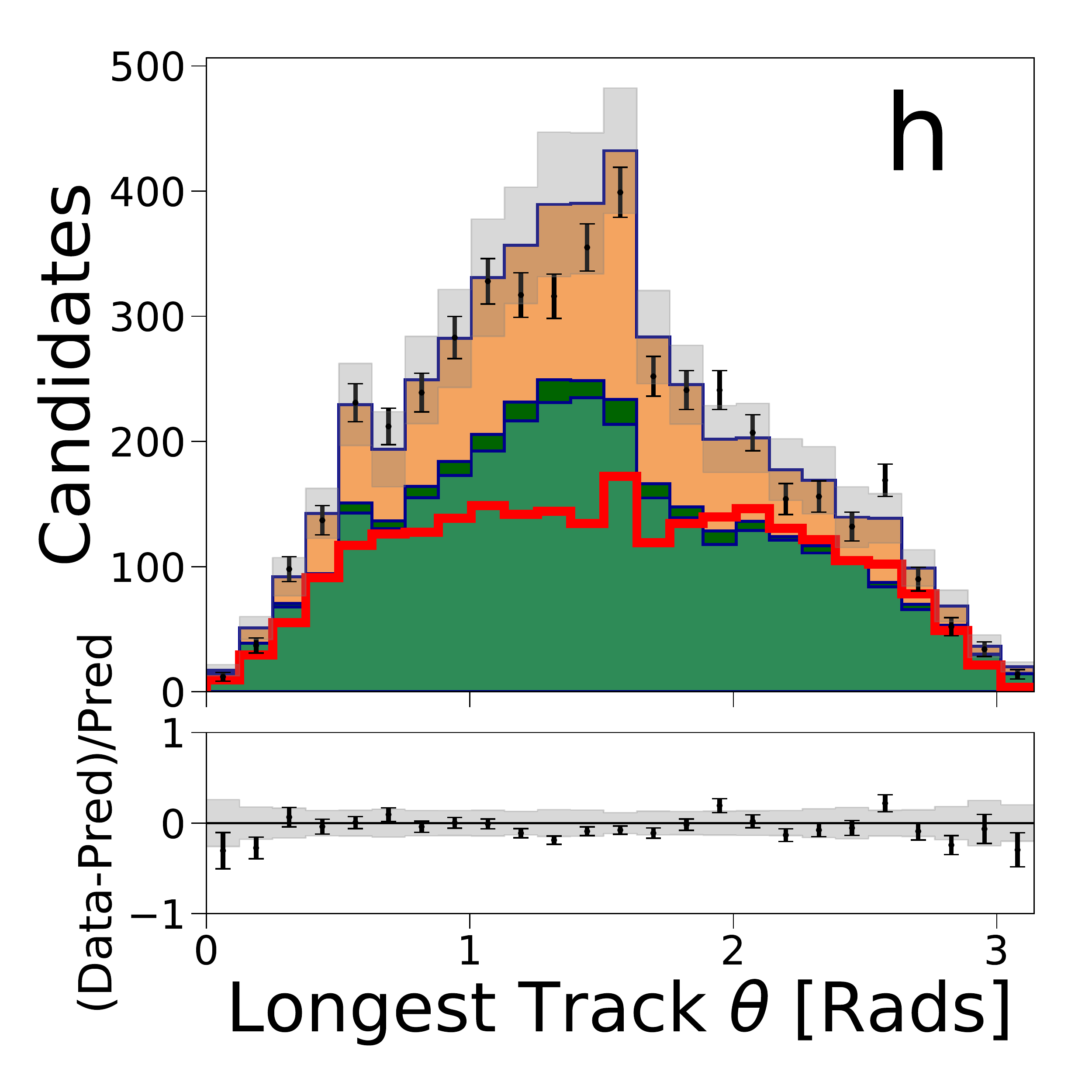}
     \includegraphics[width=0.325\textwidth]{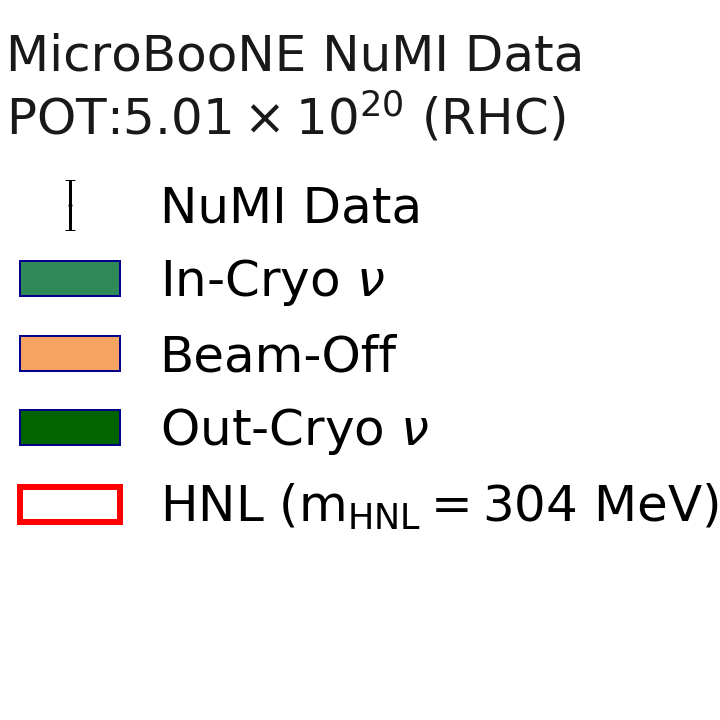}
      \caption{Sum of the background predictions (beam-off, in-cryostat $\nu$, out-of-cryostat $\nu$) for several BDT input variables:
      (a) slice energy $E_{\textrm{sl}}$,
      (b) candidate angle $\beta$,
      (c) candidate opening angle $\alpha$,
      (d) candidate mass $m_{\textrm{LLP}}$,
      (e) length $\ell_t$ of longer track,
      (f) length $\ell_t$ of shorter track,
      (g) angle $\phi$, and (h) angle $\theta$ of the longest track in the MicroBooNE coordinate system. The overlaid variable distributions are for an example mass of $\mhnl=304$~MeV. An arbitrary signal normalisation is used
      for visibility. The distributions are shown for 
      Run~3. The grey bands represent the combined statistical and systematic uncertainty on the prediction. }
  \label{fig:BDTinputdists}
\end{figure*}

\section{Boosted Decision Tree}
We use the \texttt{XGBoost} gradient boosting library~\cite{Chen:2016:XST:2939672.2939785} 
to train Boosted Decision Trees (BDTs) that discriminate between the
LLP signal and the background candidates passing the initial selection. 
A separate BDT is trained for each $m_{\rm HNL}$ and $m_{\rm HPS}$ mass point.
The BDTs are trained with $30\%$ of 
the selected in-cryostat $\nu$ sample and $50\%$ of each of the generated signal samples. 
To improve performance of the BDT training, we require that $>90\%$ of the hits in the slice are created by the LLP decay products. This procedure excludes mis-reconstructed signal events.

\begin{figure*}[htbp]
  \centering
\includegraphics[width=0.96\textwidth]{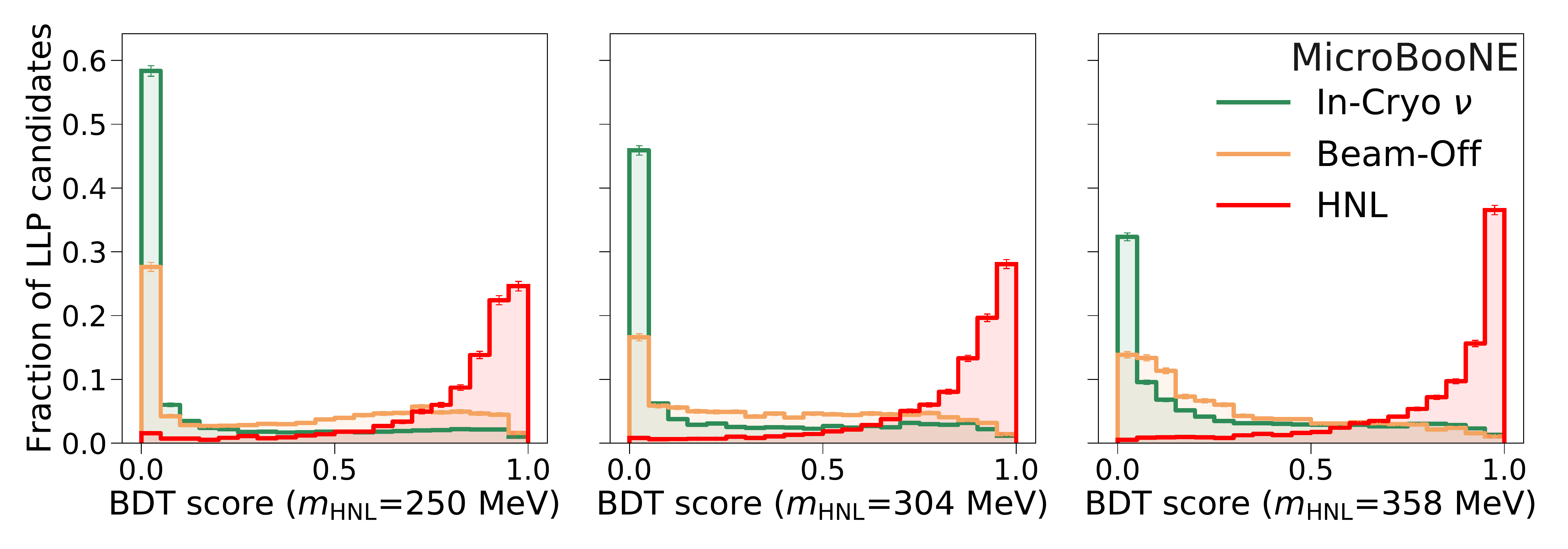}
\includegraphics[width=0.96\textwidth]{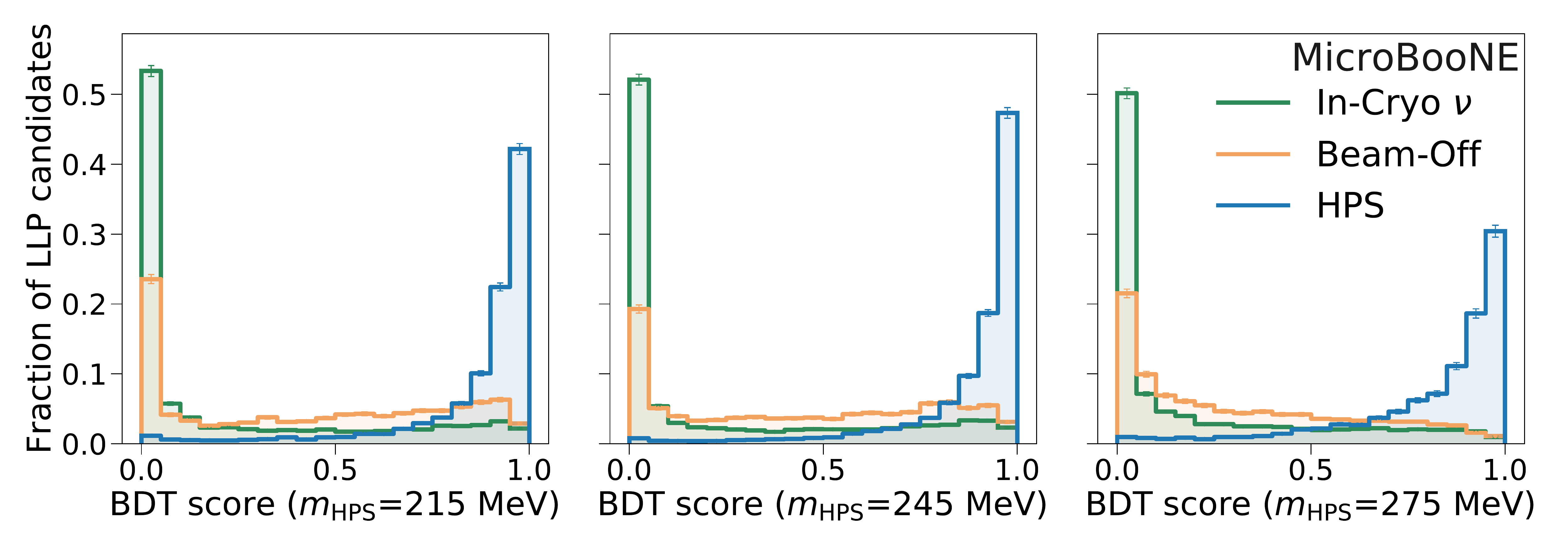}
  \caption{The Run~3 (RHC) BDT performance on the two main background samples and on representative HPS and HNL mass points.}
  \label{fig:BDTperf}
\end{figure*}

The background sample contains events where hits from overlaid cosmic events are mis-reconstructed as signal candidates. 
Therefore, cosmic-ray background is rejected by the BDT, even without training on a beam-off sample.
In total, we use 21 BDT input variables:
\begin{itemize}
    \item slice energy $E_{\textrm{sl}}$,
    \item topological score,
    \item maximum extent of the slice, $\max{(i)}$, and minimum extent, $\min{(i)}$, with $i=x,y,z$,
    \item multiplicities $N_{\rm tot}$, $N_{\rm sh}$, and $N_{\rm tr}$,
    \item candidate angle $\beta$,
    \item candidate opening angle $\alpha$,
    \item candidate mass $m_{\textrm{LLP}}$,
    \item angles $\theta$ and $\phi$ of the longest track, as defined in the MicroBooNE coordinate system, and
    \item length, the PID score $S_{\rm PID}$ and the track score of the two tracks forming the candidate.
\end{itemize}

To compare data and background simulation, and to demonstrate the sensitivity to an LLP signal, we show eight of the more important variables in the BDT in Fig.~\ref{fig:BDTinputdists}. In general, we observe good agreement between the background prediction and the data across all variables, both in shape and normalization. The distributions shown are for Run~3 (RHC). The distributions for Run~1 (FHC) are similar with an increased cosmic-ray contribution (see Table~\ref{tab:presel}) as the CRT was not yet operational. 

In Fig.~\ref{fig:BDTinputdists}, we overlay the signal distribution for a typical mass point, at $\mhnl=304$~MeV. As expected, the average reconstructed invariant mass is centered around this value. The slice energy peaks at $E_{\textrm{sl}}\approx 105$~MeV, about $20$~MeV above the expected value from Eq.~\ref{eq:Esl}. This shift is related to the ionization energy deposited by Michel electrons. The slice energy is the variable with the best sensitivity to signal in the BDTs. The candidate opening angle is expected to peak at $\cos\alpha=-0.34$ for this mass. 
The candidate angle $\beta$ with respect to the direction from the absorber is peaked at $\cos{\beta}=1$ for a large fraction of the HNL signal candidates. 

The lengths of the tracks are measures 
of the momenta of the two particles, 
which depend on the decay kinematics. 
The direction of the longer decay track is included as the angles $\theta$ and $\phi$ of the track in the detector coordinate system. The direction (in radians) from the absorber corresponds to $\theta=2.20$ and $\phi =1.15$. The angle $\phi$ helps to reject cosmic rays, which are aligned with $\phi=\pm \pi/2$.

\begin{figure*}[htbp]
\centering
\begin{picture}(160,160)
\put(-165,0){\includegraphics[width=0.48\textwidth]{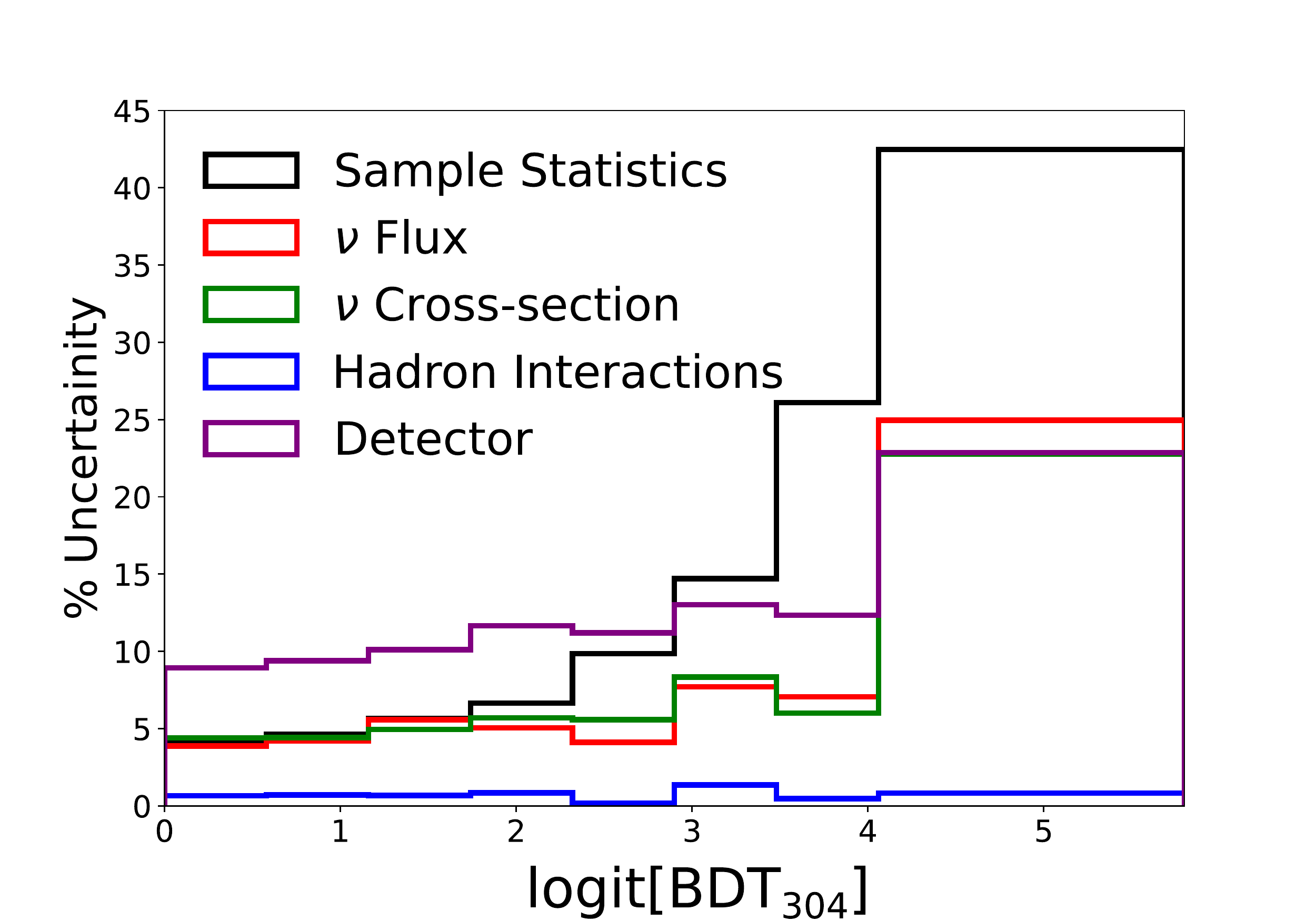}}
\put(95,0){\includegraphics[width=0.48\textwidth]{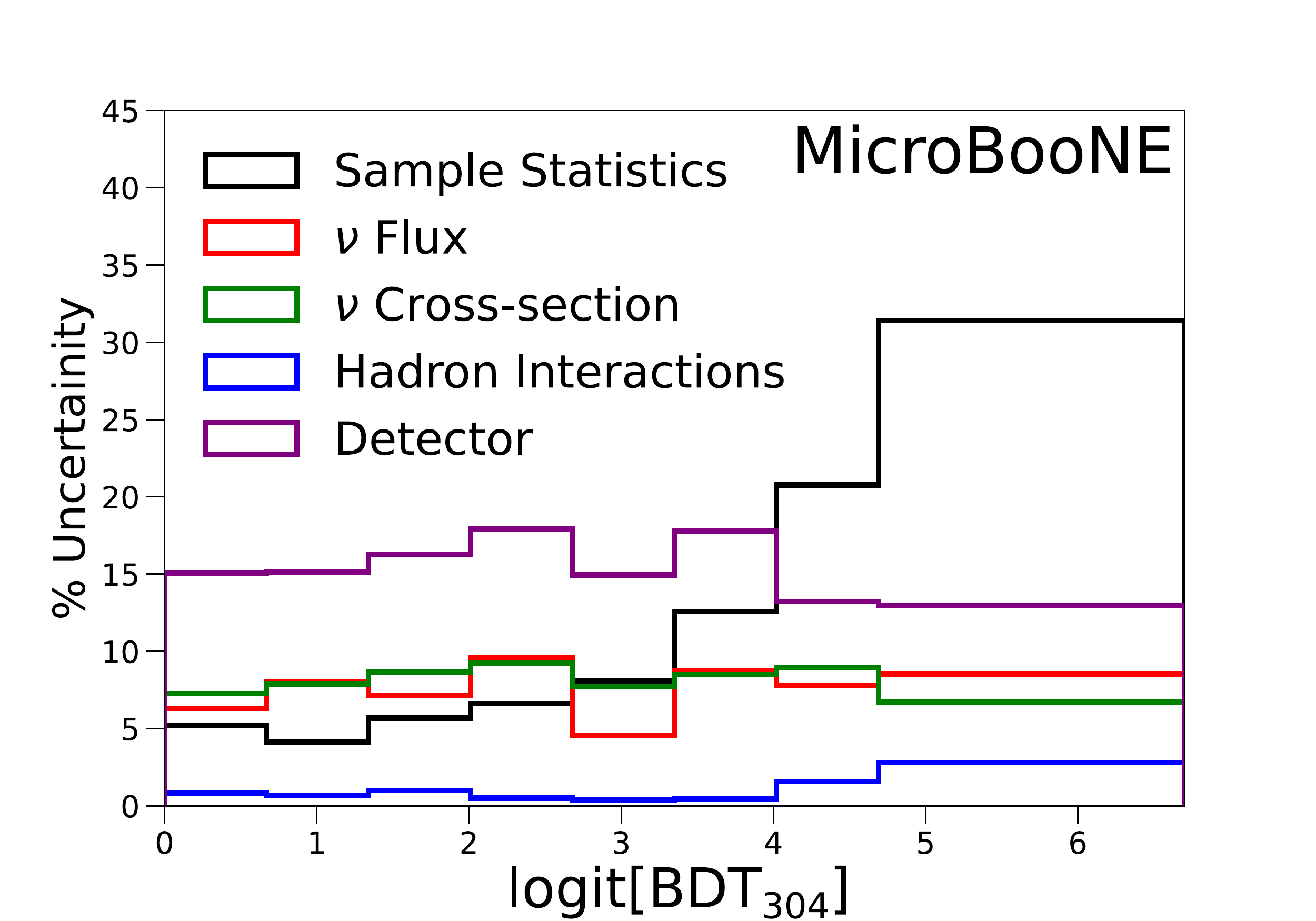}}
\put(-125,71){\Large (a)}
\put(137,71){\Large (b)}
\end{picture}
  \caption{Fractional uncertainty on the background prediction for the (a) Run 1 (FHC) and (b) Run 3 (RHC) samples as a function of the BDT score, shown separately for the main sources of uncertainty. The BDT has been trained for a signal mass of $\mhnl=304$ MeV. }
  \label{fig:BDTsys}
\end{figure*}

Figure~\ref{fig:BDTperf} shows the distributions of the BDT scores trained for three different HNL and HPS masses.
The BDTs offer strong rejection against background candidates from neutrino interactions.
Background candidates from mis-reconstructed cosmic-ray tracks, as found in the beam off sample, 
are also rejected with high efficiency.
The discrimination between signal and background improves slightly with LLP mass due to the higher energy of the decay particles, leading to improved kinematic reconstruction.

In Fig.~\ref{fig:BDTperf}, the full range of BDT scores lies in the range [0,1]. The final score distribution
is shown after a transformation using the inverse of the logistic function has been applied, which maps the score to a range [$-\infty$,$\infty$]. We only consider candidates with BDT score $>0$ after the transformation, corresponding to $>0.5$ in the original distribution, since this region contains $>90\%$ of the signal and hence dominates the sensitivity.

\section{Systematic Uncertainties}
\label{sec:systematics}

Uncertainty sources are considered for the background samples and signal samples by applying variations that modify the BDT score distributions.
For the simulated in-cryostat $\nu$ background sample, we consider the impact of the flux simulation, cross-section modeling, hadron interactions with argon, and detector variations:

\begin{itemize}
    \item {\bf Flux Simulation.}
Uncertainties on the neutrino flux arise primarily from the rates and kinematics of hadron production in the beamline. 
The \texttt{PPFX} package is used to estimate these uncertainties.
Each of the parameters in the constrained flux prediction is simultaneously sampled within its estimated uncertainties to produce alternative flux predictions. 
Flux uncertainties also include variations in the beamline conditions, e.g., changes in horn current, horn position, and beam spot location. However, these beamline condition uncertainties are neglected, as they have been shown to be small~\cite{MicroBooNENuMInuexsec1,MicroBooNENuMInuexsec2}
and were therefore not reassessed. 

\item {\bf Cross-section Modeling.}
We assess the uncertainties due to the modeling of neutrino cross-sections by varying 44 of the parameters used by the \texttt{GENIE} generator. A full discussion can be found in Ref.~\cite{MicroBooNEGenieTune}.

\item {\bf Hadron Interactions.}
Hadrons produced in neutrino interactions interact strongly, affecting their propagation through argon. The description of hadron interactions therefore impacts the event reconstruction and the description of the neutrino background.
These uncertainties are assessed using \texttt{GEANT4Reweight}~\cite{GEANT4Reweight} by considering variations in the \texttt{GEANT4} cross-section model for charged pions and protons. 

\item {\bf Detector Modeling.}
Uncertainties arising from detector modeling are estimated by re-simulating the detector configuration using the same generated input event sample. The variations include effects due to the light simulation, reduction in the light yield, increase of the Rayleigh scattering length, and attenuation of the light in the argon. 

Uncertainties related to charge reconstruction are assessed using data-driven modifications of the the waveforms on the TPC wires, as discussed in Ref.~\cite{WireModPub}. Additional variations due to the space charge mapping and ion recombination model are simulated and assessed separately.
\end{itemize}

We expect only a small number of
background events in the signal region of the BDT distribution due to the high purity of the event selection. We therefore extrapolate
detector-modeling uncertainties from higher statistics regions of the 
distribution to the signal region, assuming they are constant.

The beam-off sample is taken from data and therefore has no associated systematic uncertainties other than the statistical fluctuations in the sample. The contribution of the out-of-cryostat sample to the final sample is small. The out-of-cryostat sample normalization (see Section~\ref{sec:data}) is set as an unconstrained parameter in the final fits and found to be negligible.

The impact of the systematic uncertainties on the BDT score distribution is shown in Fig.~\ref{fig:BDTsys} for the Run 1 (FHC) and Run 3 (RHC) background samples, with the BDT trained for a signal of $\mhnl=304$~MeV. 
The fractional uncertainties are given relative to the
total background prediction, which is the sum of beam-off, in-cryostat $\nu$, and out-of-cryostat $\nu$ samples.

The dominant uncertainty in the signal region at high BDT scores is due to the statistical uncertainty of the background samples, which is
a consequence of the high purity of the signal selection. 
Detector modeling uncertainties are $\approx (10-20)\%$, neutrino flux and cross-section uncertainties are each $\approx (5-10)\%$ for most bins, while all other uncertainties are negligible. The systematic uncertainties are separately evaluated for all signal training points (signal masses and FHC/RHC), with consistent results. As expected, the sensitivity of the final result is dominated by the the RHC data set.

The dominant contribution to the systematic uncertainty on the signal sample arises from uncertainties on the rate of kaon production at rest in the NuMI absorber. This normalisation uncertainty is taken from the evaluation of the MiniBooNE collaboration to be $\pm 30\%$~\cite{MiniBooNEKDAR}, as discussed in Section~\ref{sec:flux}.

The hadron interaction uncertainty will mainly affect the modeling of pions produced in the HNL decay. The relative uncertainties on the BDT distribution are $\approx 2\%$ for hadron interaction modeling and $<15\%$ for detector modeling. The longest track is assigned to be the muon in HNL decays; the systematic uncertainty due to this choice is also negligble. 

Finally, a $2\%$ uncertainty on the number of POT delivered is estimated from the uncertainty on the beamline toroidal monitor measurement~\cite{Aliaga:2016oaz}, which is taken as fully correlated uncertainty for all samples.
 
\section{Results}
\label{sec:results}
\begin{figure*}[ht!]
  \centering
  \includegraphics[width=0.97\textwidth]{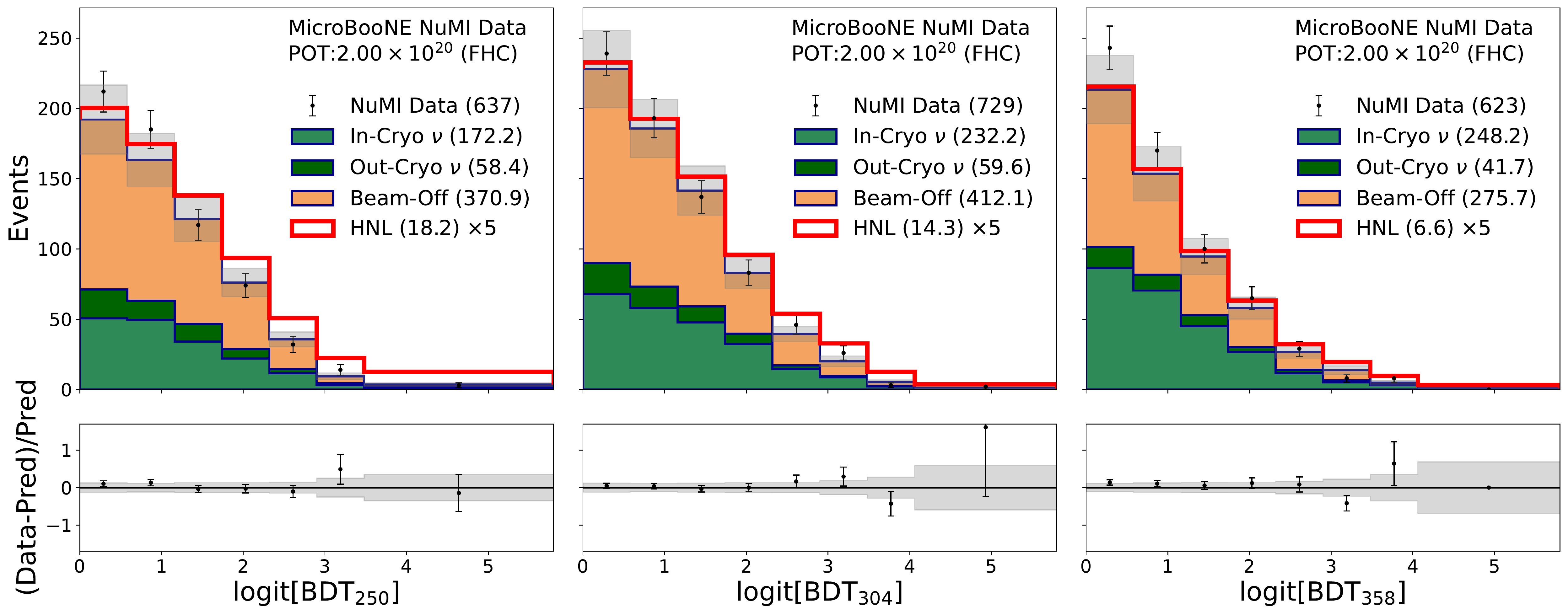}
    \includegraphics[width=0.97\textwidth]{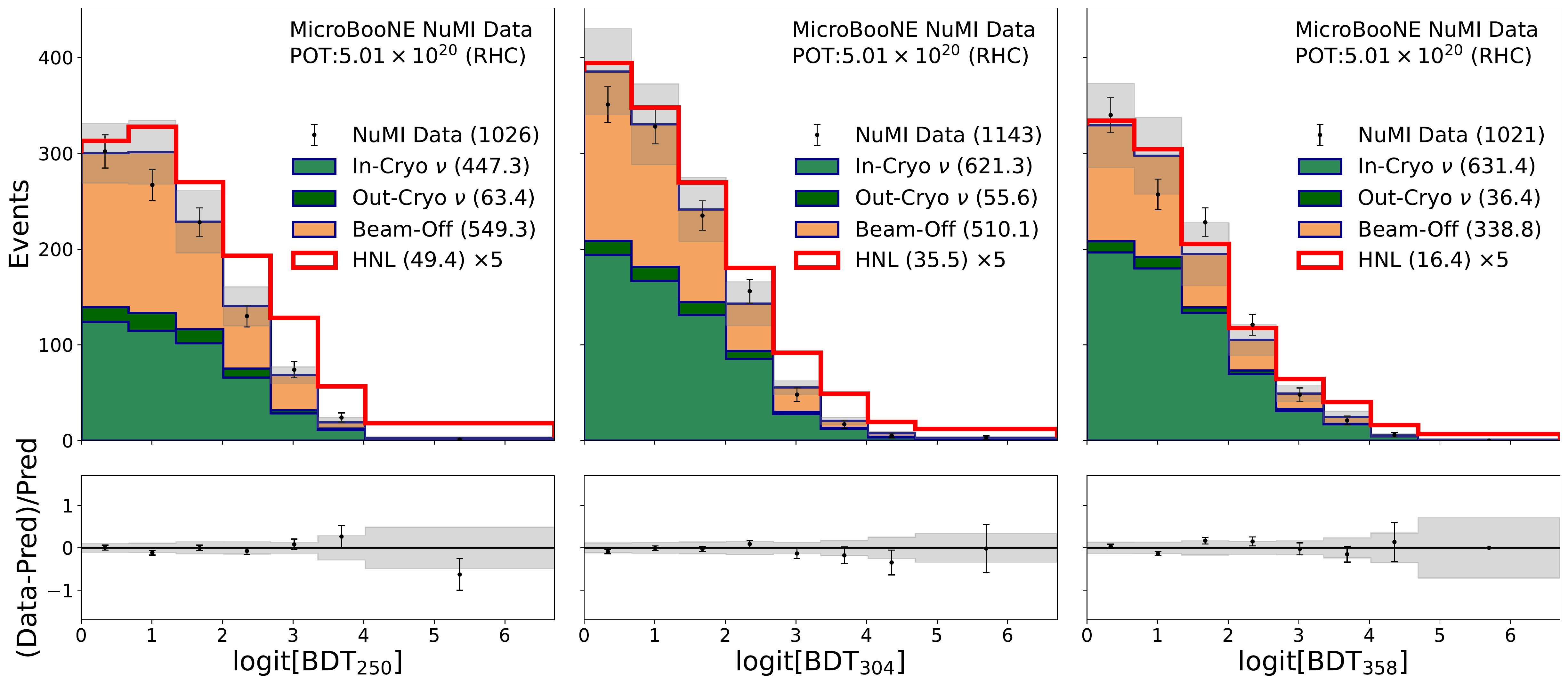}
\caption{
Sum of the background predictions (beam-off, in-cryostat $\nu$, out-of-cryostat $\nu$) for the BDT score distribution compared to data and to the expected
signal distribution for $\mhnl=250$~MeV, $304$~MeV, and $358$~MeV.
 The HNL signal distributions are normalized to the number of events excluded at the $90\%$~CL (quoted in brackets on the plot), scaled up by a factor of $5$ for better visibility and added to the background distribution. The top row shows Run 1 (FHC) and the bottom row Run 3 (RHC) data.
 The bands represent the combined statistical and systematic uncertainty on the prediction.}
  \label{fig:BDTdists1}
\end{figure*}

\begin{figure*}[ht!]
  \centering
      \includegraphics[width=0.97\textwidth]{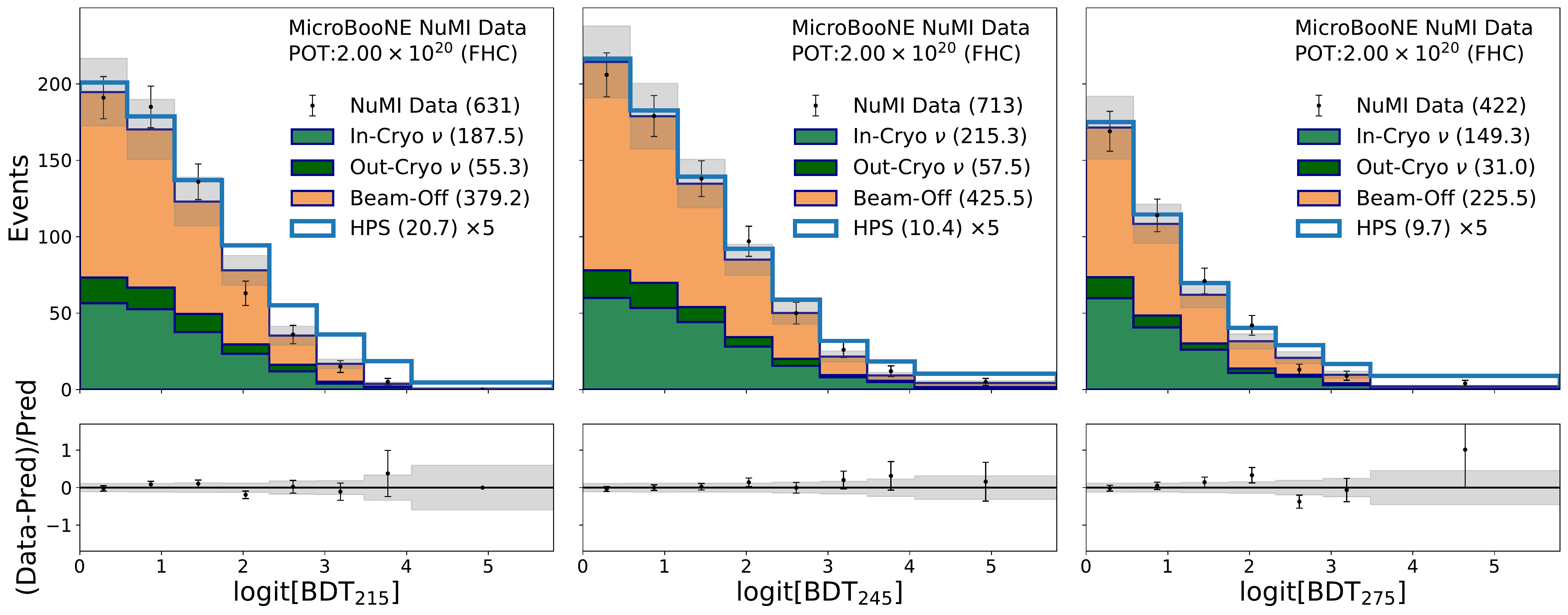}
        \includegraphics[width=0.97\textwidth]{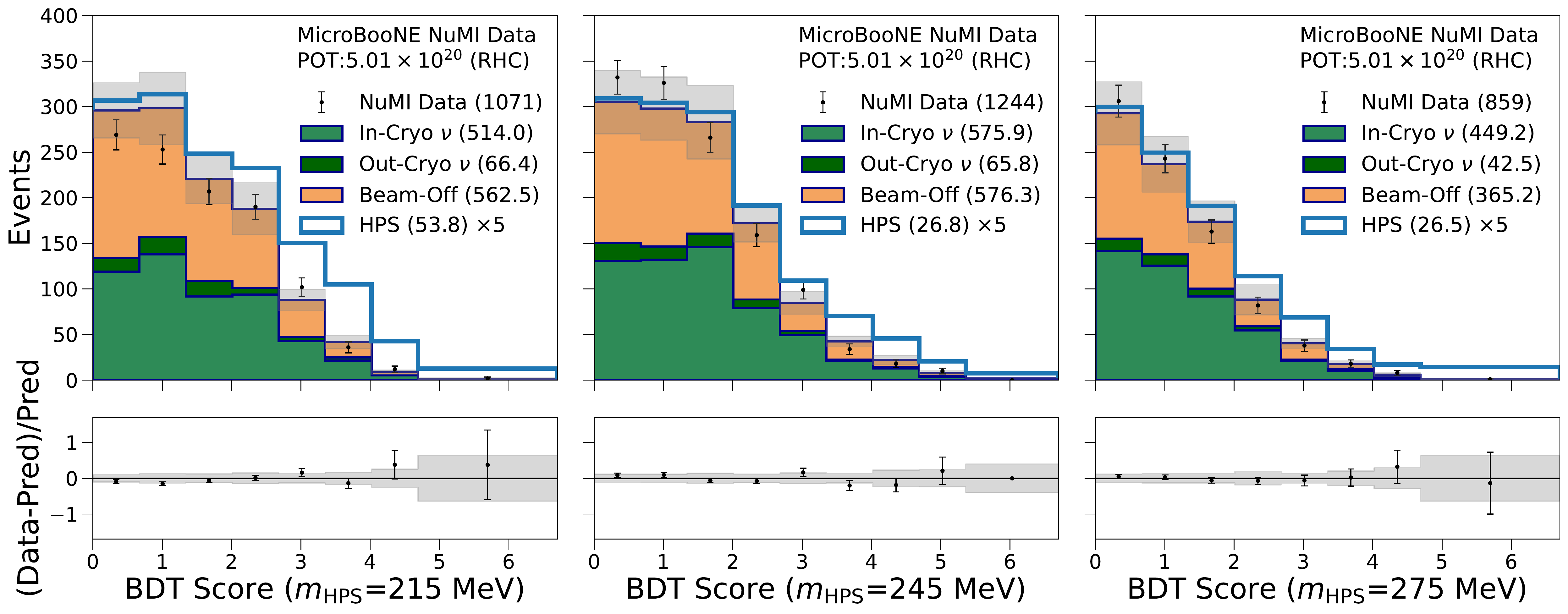}
\caption{
Sum of the background predictions (beam-off, in-cryostat $\nu$, out-of-cryostat $\nu$) for the BDT score distribution compared to data and to the expected
signal distribution for $\mhps=215$~MeV, $245$~MeV, and $275$~MeV.
 The HPS signal distributions are normalized to the number of events excluded at the $90\%$~CL (quoted in brackets on the plot), scaled up by a factor of $5$ for visibility and added to the background distribution. The top row shows Run~1 (FHC) and the bottom row Run~3 (RHC) data.
 The bands represent the combined statistical and systematic uncertainty on the prediction.}
  \label{fig:BDTdists2}
\end{figure*}

We apply the BDTs trained for each signal mass point to the data and the simulation. In Figs.~\ref{fig:BDTdists1} and \ref{fig:BDTdists2} we show a comparison of the resulting BDT distributions for the data to the sum of the background predictions, using a selection of representative signal mass points.
In events with more than one candidate, we retain only the candidate with the most signal-like BDT score. The background predictions and the data are in good agreement across both of the run periods studied. 
%Since no significant excess consistent with LLP decays is observed, we proceed to set limits.

The BDT score distributions are used as input to a modified frequentist CL$_s$ calculation~\cite{Junk:1999kv,Read:2002hq} to set upper limits on the signal strength for each model and mass point. Test statistics are constructed from the log-likelihood ratio (LLR) of the signal-plus-background (S+B) and background-only (B) hypothesis for each LLP mass.
The BDT distributions for each run period (FHC and RHC) enter the limit setting as separate channels before their likelihoods are combined.

The systematic uncertainties on the background and signal predictions are taken into account using Gaussian priors.
The signal and background predictions are separately fitted to the data  distributions under the background-only and signal-plus-background hypotheses.
The systematic uncertainties are allowed to vary within their defined priors to maximize the respective likelihood functions. This reduces the effect of the systematic uncertainties on the sensitivity of the search~\cite{Fisher:2006zz}.

The background predictions are fitted separately for the two 
run periods, with uncorrelated systematic uncertainties. 
We find that this assumption about the correlations 
has negligible impact on the result.
The flux systematic uncertainties on the signal are taken to be
correlated between the two data periods. 

The confidence levels are calculated by integrating the LLR distributions, 
which are derived using pseudo-experiments, under either
the S+B (CL$_{s+b}$) or the B-only hypotheses (CL$_b$).
The excluded signal rate is defined as the scaling of the signal strength for which the confidence level for signal reaches ${\rm CL}_s = {\rm CL}_{s+b}/{\rm CL}_b=1-0.9$.

\begin{figure*}[ht!]
  \centering
  \includegraphics[width=1\textwidth]{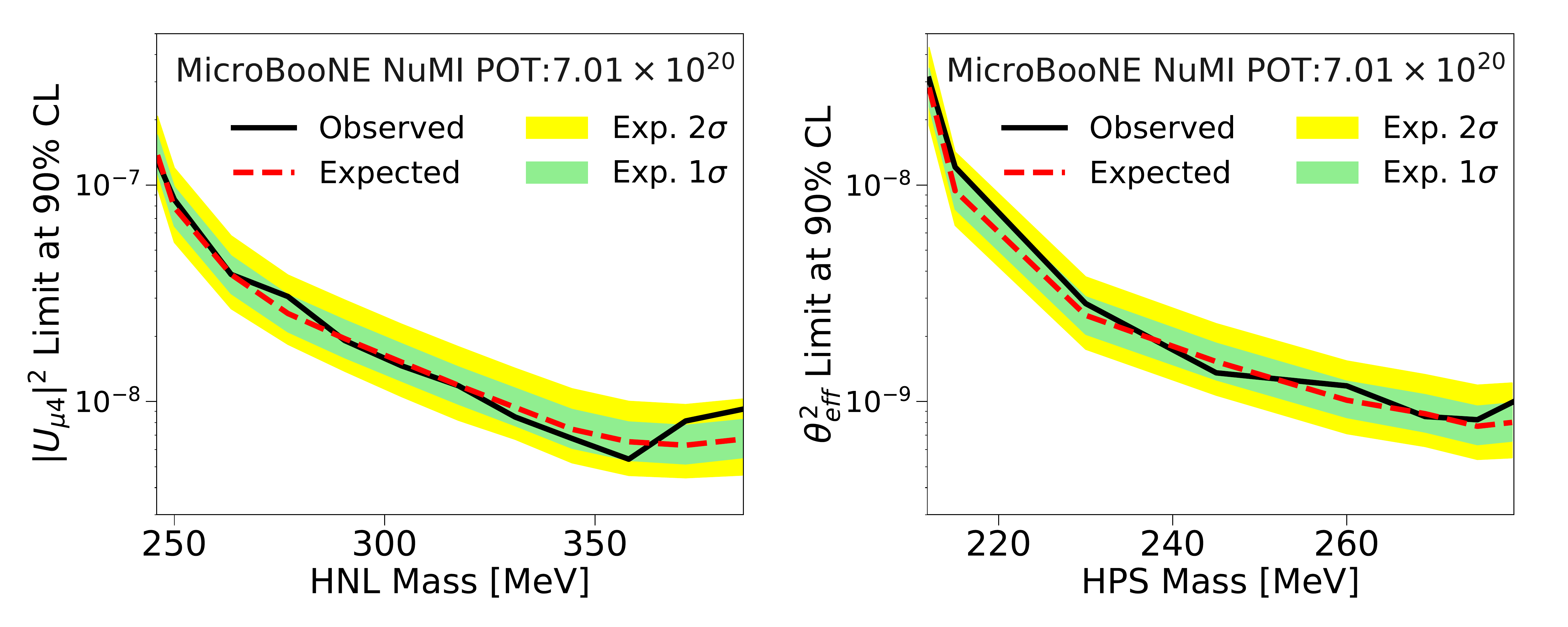}
\caption{
Limits at the $90\%$ confidence level as function of mass for (a) $\mumix$, assuming a Majorana HNL decaying into $\mu\pi$ pairs, and (b) $\thetaeffmix$ of an HPS decaying into $\mu\mu$ pairs. 
The observed limits are compared to the median expected limit with the $1$ and $2$ standard deviations ($\sigma$) bands.}
  \label{fig:brazil}
\end{figure*}

\begin{table}[ht!]
\centering
\setlength{\tabcolsep}{10pt} 
\renewcommand{\arraystretch}{1.05}
\caption{The $90\%$~CL observed and median expected limits on $\mumix$ as
a function of $\mhnl$ for a Majorana HNL.}
\begin{tabular}{cccc}
   \hline  \hline
   $m_{\rm HNL}$  & \multicolumn{3}{c}{limit on $\mumix$ ($\times 10^{-8}$)}\\
 (MeV)& observed &  median & $1$ s.d. range 
              \\
            
\hline
$246.0$ & 12.9 & 13.7 & 11.3--17.0 \\
$250.0$ & 8.57 & 7.89 & 6.46--9.83 \\
$263.5$ & 3.86 & 3.86 & 3.15--4.71 \\
$277.0$ & 3.05 & 2.55 & 2.10--3.11\\
$290.5$ & 1.91 & 1.95 & 1.59--2.38\\
$304.0$ & 1.46 & 1.52 & 1.24--1.85\\
$317.5$ & 1.18 & 1.19 & 0.97--1.45\\
$331.0$ & 0.85 & 0.94 & 0.77--1.15\\
$344.5$ & 0.67 & 0.74 & 0.61--0.92\\
$358.0$ & 0.54 & 0.65 & 0.53--0.80\\
$371.5$ & 0.81 & 0.63 & 0.51--0.78\\
$385.0$ & 0.92 & 0.67 & 0.55--0.83\\
\hline\hline
\end{tabular}
\label{tab:HNLResults}
\end{table}

The observed and median expected $90\%$~CL limits on $\mumix$ are shown for each HNL mass point in Table~\ref{tab:HNLResults} and in Fig.~\ref{fig:brazil}, and the corresponding limits on $\thetaeffmix$ for the HPS model in~Table~\ref{tab:HPSLimits}. 
The $1$- and $2$-standard deviation intervals cover the range of expected limits produced by $68\%$ and $95\%$ of background prediction outcomes around the median expected value.
The observed limits are contained in the $1$-standard-deviation interval for all mass points with the exception of $\mhnl=371.5$~MeV and $385.0$~MeV, and for $\mhps=215$~MeV, where the observed limit lies within $2$ standard deviations.
We use a linear interpolation between the mass points when drawing contours. 

We derive the HNL limits assuming that HNLs are Majorana particles. For a Dirac HNL, only decays to the charge conjugated final state $\mu^-\pi^+$ are allowed in $K^+$ decays. The expected number of decays is therefore a factor of two smaller for the same $\mumix$ value.
The limits for Dirac HNLs are calculated from the Majorana limit by applying a factor of $\sqrt{2}$ to account for the reduced decay rate, since the difference due to angular distributions of the decay is found to be negligible.

\begin{table}[htb]
\centering
\setlength{\tabcolsep}{5pt} 
\renewcommand{\arraystretch}{1.05}
\caption{The $90\%$ CL observed and expected limits on $\thetaeffmix$ obtained by this analysis, and the 
$\theta^2$ contour derived from the
$\thetaeffmix$ limits.}
\begin{tabular}{c|ccc|cc}
    \hline\hline
  $\mhps$     & \multicolumn{3}{c|}{limit on $\thetaeffmix$ ($\times 10^{-9}$)} & \multicolumn{2}{c}{$\theta^2$ range ($\times 10^{-9}$)} \\
   (MeV)        & \multicolumn{1}{c}{obs.} &  \multicolumn{1}{c}{median} & \multicolumn{1}{c|}{$1$~s.d.} & low & high\\
%& \multicolumn{3}{c}{$\times 10^{-9}$} &
%\multicolumn{2}{|c|}{$\times 10^{-9}$} 
%              \\
\hline
$212$ & $30.83$ & $28.25$ & $23.07$--$34.68$ & $\phantom{0}31.3\phantom{0}$ 
& $25000$  \\
$215$ & $12.1$ & $9.41$ & $7.73$--$11.5$ &$12.5$ & $1490$ \\
$230$ & $2.83$ & $2.50$ & $2.04$--$3.05$ & $3.14$ & $90.0$ \\
$245$ & $1.36$ & $1.53$ & $1.26$--$1.86$ & $1.63$ & $32.0$ \\
$260$ & $1.18$ & $1.01$ & $0.84$--$1.24$ & $1.55$ & $13.2$\\
$269$ & $0.85$ & $0.88$ & $0.72$--$1.08$ & $1.14$ & $10.2$\\
$275$ & $0.82$ & $0.77$ & $0.63$--$0.95$ & $1.09$ & $5.05$\\
$279$ & $0.99$ & $0.80$ & $0.66$--$0.99$ &  \multicolumn{2}{c}{N.A.}  \\
\hline\hline
\end{tabular}
\label{tab:HPSLimits}
\end{table}

Table~\ref{tab:HPSLimits} shows the values of $\thetamix$ that correspond to the lower and upper bounds of the excluded region for each mass \mhps, where the upper boundary is due to the short lifetime of the HPS.
For the highest mass point at $\mhps=279$~MeV,
the number of HPS reaching the detector before decaying is too small to derive a limit for any of the $\theta^2$ values within the excluded $\thetaeffmix$ range.

\begin{figure*}[htbp]
  \centering
  \includegraphics[width=0.8\textwidth]{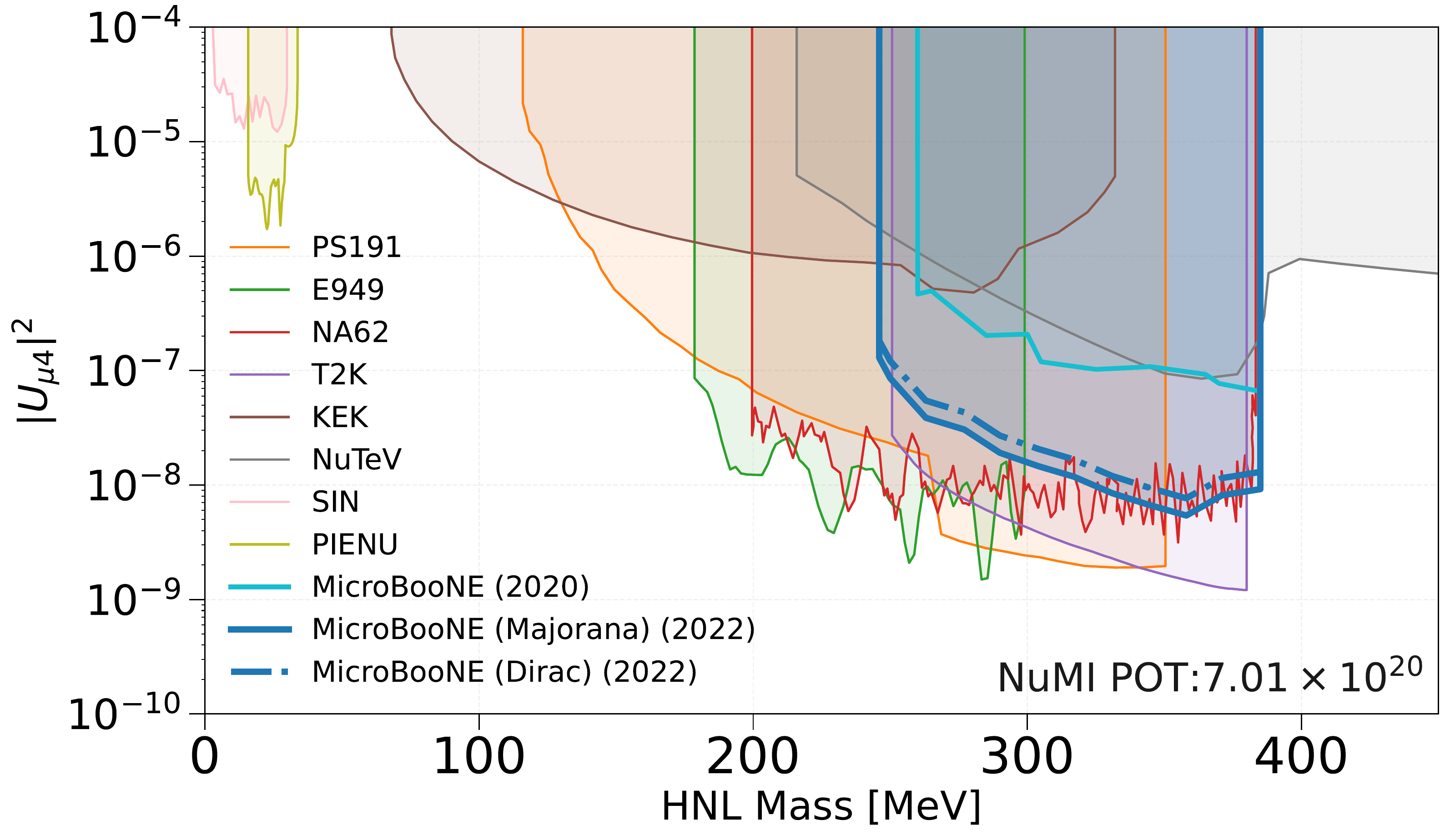}
  \caption{Limits on $\mumix$ at the $90\%$ CL as function of mass for Majorana and Dirac HNL compared to the
  results of the SIN~\cite{Daum:1987bg}, PIENU~\cite{PIENU:2019usb}, KEK~\cite{Hayano:1982wu}, NA62~\protect\cite{NA62:2021bji}, E949~\cite{Artamonov:2014urb}, PS191~\protect\cite{Bernardi:1987ek}, T2K~\protect\cite{Abe:2019kgx} and NuTeV~\cite{Vaitaitis:1999wq} collaborations.}
  \label{fig:limscomphnl}
\end{figure*}

\begin{figure*}[htbp]
  \centering
  \includegraphics[width=0.8\textwidth]{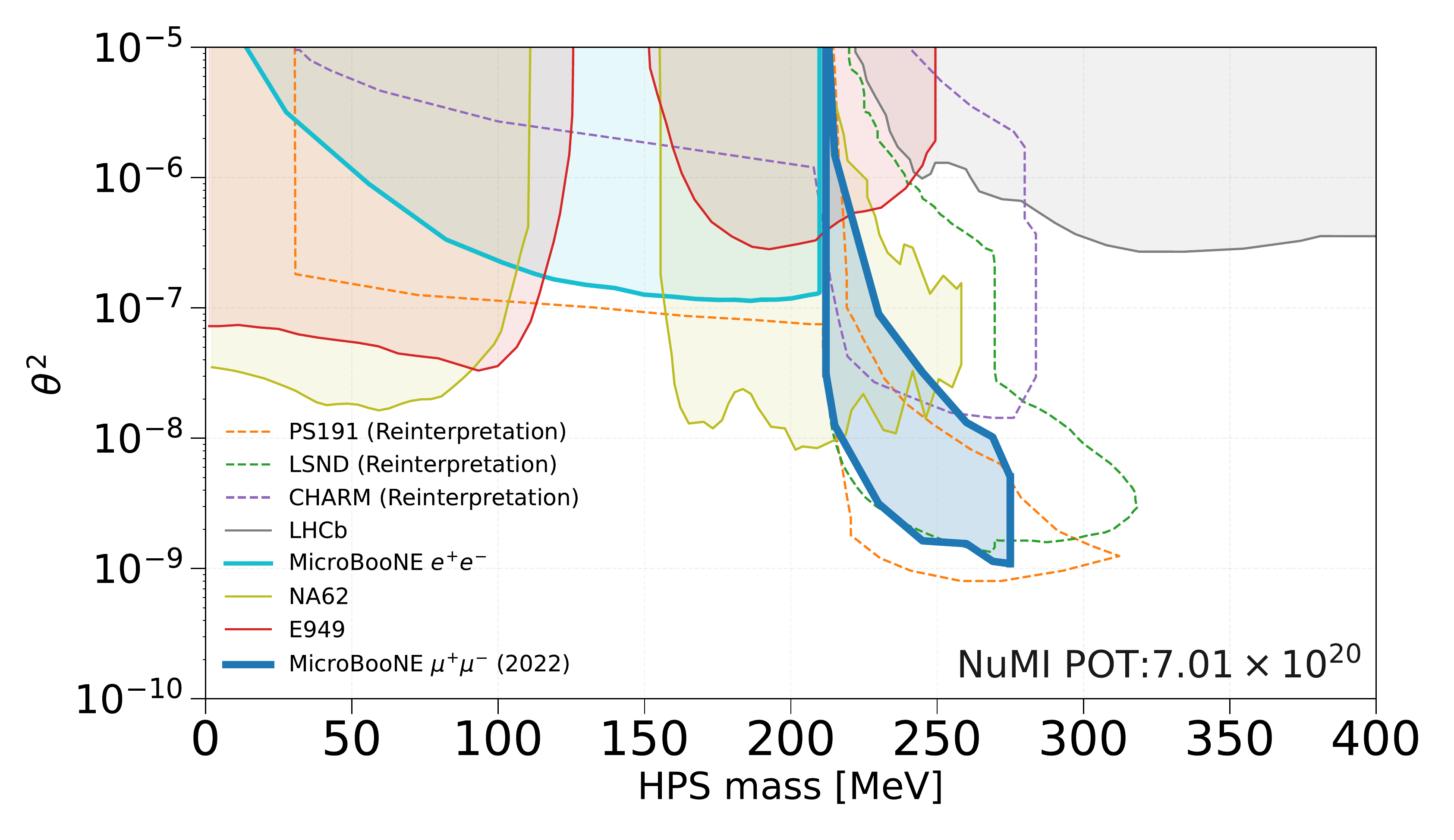}
  \caption{Limits at the $90\%$~CL on the scalar-Higgs mixing angle $\theta$ as a function of $\mhps$ compared to reinterpretations of 
  CHARM~\protect\cite{Winkler:2018qyg}, LSND~\protect\cite{LSNDReint}, and PS191~\protect\cite{PS191_reint} measurements.
  In other mass ranges, limits are from a MicroBooNE search for the $e^+e^-$ final state~\protect\cite{UbooneHPSPaper} (at the $95\%$~CL), and from searches by the NA62~\protect\cite{NA62:2021bji,NA62:2021zjw} and E949 collaborations~\protect\cite{BNL-E949:2009dza_HPS} for charged kaon decays to pions and an HPS.
The LHCb collaboration performed two searches for an HPS with short lifetime, which would be produced and subsequently decay within the detector~\protect\cite{LHCb:2015nkv,LHCb:2016awg}. The joint
coverage of the LHCb result is shown at the $95\%$~CL.
  }
  \label{fig:limscomphps}
\end{figure*}

\section{Comparison with Existing Limits}

In Figs.~\ref{fig:limscomphnl} and \ref{fig:limscomphps}, we compare the observed limits to the existing experimental limits in similar regions of parameter space for both models. The results extend MicroBooNE's sensitivity to $\mumix$ by approximately an order of magnitude compared to the previous MicroBooNE HNL result~\cite{Abratenko:2019kez}.

Like the MicroBooNE detector, the T2K~\cite{Abe:2019kgx} and NuTeV detectors~\cite{Vaitaitis:1999wq} were located in a neutrino beamline. The PS191 experiment~\cite{Bernardi:1985ny,Bernardi:1987ek} at CERN was specifically designed to search for massive decaying neutrinos. The NA62~\cite{CortinaGil:2017mqf} and E949~\cite{Artamonov:2014urb} collaborations performed a peak search for HNLs in kaon decays. The muon spectrum measured in stopped $K^+\to\mu^+\nu$ decays $(K_{2\mu})$ has also been used to set limits on HNLs~\cite{Asano:1981he,Hayano:1982wu}.

In the mass range $300<\mhnl<385$~MeV, this search has similar sensitivity as the NA62 experiment~\cite{NA62:2021bji}. 
The E949~\cite{Artamonov:2014urb}, PS191~\cite{Bernardi:1987ek} and T2K~\cite{Abe:2019kgx} limits are stronger across the range $300<\mhnl<385$ MeV.
The T2K collaboration provides no limit point for masses above $380$~MeV. Here, the MicroBooNE limit is of equal or greater sensitivity than the NA62 result.

For the HPS model, we constrain a region of parameter space for $212<\mhps<275$~MeV not previously excluded by any dedicated experimental search.
The existing limits in this region are reinterpretations of decades old CHARM~\cite{Winkler:2018qyg}, LSND~\cite{LSNDReint}, and PS191~\cite{PS191_reint} measurements, performed by authors outside the respective collaborations
without access to the original experimental data or MC simulation.
 Reinterpretations depend on external beamline, flux, and detector simulations. 
 If the signal topology differs from the original selection criteria, the results also depend on estimated detection efficiencies. In the case of the CHARM experiment, 
 the more recent sensitivity estimate in Ref.~\cite{Winkler:2018qyg} disagrees by nearly an order of magnitude from the estimate in Ref.~\cite{Clarke:2013aya}.
 %Only one reinterpretation has been published for the PS191~\cite{PS191_reint} and LSND~\cite{LSNDReint} experiments, respectively.
 
\section{Summary} \label{sec:summary}
We present a search for long-lived particles (LLP) using NuMI beam data corresponding to $7.01\times10^{20}$ POT recorded with the MicroBooNE detector. The results are interpreted within two models, where the LLP is either
a heavy neutral lepton (HNL) or a Higgs portal scalar (HPS). The LLPs are assumed to be produced by $K^+$ mesons decaying at rest in the NuMI absorber. The signature in the MicroBooNE liquid-argon detector are HNL decays 
into $\mu^{\pm}\pi^{\mp}$ pairs or HPS decays into $\mu^+\mu^-$ pairs.

The main sources of background are neutrino and cosmic-ray interactions, where the majority of the neutrino events
are from charged-current muon neutrino interactions. 
To reject background, we select data recorded in-time with the NuMI beam and consistent with the LLP signatures in the liquid argon. 
The LLPs originating in the NuMI absorber
enter the detector from a different direction than the majority of beam neutrinos. The decay products also have a fixed energy for a given LLP mass. These kinematic properties are used to discriminate signal from background. 
The combined reconstruction and selection efficiencies for LLP decays 
are between $13\%$ and $30\%$, increasing with LLP mass.
To further improve discrimination between signal and background, we train and apply a BDT with 21 input variables for each mass point. 

No significant excess is observed in the BDT score distributions.
In the absence of signal, we employ the modified frequentist CL$_s$ method to derive limits on the model mixing parameters $\mumix$ and $\theta^2$. All limits are presented at the $90\%$ CL.

We set upper limits on the mixing parameter $\mumix$ ranging from $\mumix=12.9\times 10^{-8}$ for Majorana HNLs with a mass of $\mhnl=246$~MeV to $\mumix=0.92 \times 10^{-8}$ for $\mhnl=385$~MeV, assuming $\lvert U_{e 4}\rvert^2 = \lvert U_{\tau 4}\rvert^2 = 0$ and HNL decays
into $\mu^\pm\pi^\mp$ pairs.
These limits on $\mumix$ are of similar sensitivity to those published by the NA62 collaboration~\cite{NA62:2021bji} and they represent an order of magnitude improvement in sensitivity compared to the previous MicroBooNE result~\cite{Abratenko:2019kez}.

We also constrain the scalar-Higgs mixing angle $\theta$ by searching for HPS decays into $\mu^+\mu^-$ final states, excluding a contour in the parameter space with lower bounds of 
$\theta^2<31.3 \times 10^{-9}$ for $\mhps=212$~GeV 
and
$\theta^2<1.09 \times 10^{-9}$ for $\mhps=275$~GeV.
These are the first constraints in this region of the $\theta^2$--$\mhps$ parameter space from a dedicated experimental search. It is also the first search in this mass range using a liquid-argon TPC.

\section*{Acknowledgements}

This document was prepared by the MicroBooNE collaboration using the resources of the Fermi National Accelerator Laboratory (Fermilab), a U.S. Department of Energy, Office of Science, HEP User Facility. Fermilab is managed by Fermi Research Alliance, LLC (FRA), acting under Contract No.\ DE-AC02-07CH11359. MicroBooNE is supported by the following: the U.S. Department of Energy, Office of Science, Offices of High Energy Physics and Nuclear Physics; the U.S. National Science Foundation; the Swiss National Science Foundation; the Science and Technology Facilities Council (STFC), part of the United Kingdom Research and Innovation; the Royal Society (United Kingdom); and The European Union’s Horizon 2020 Marie Sk\l{}odowska-Curie Actions. Additional support for the laser calibration system and cosmic ray tagger was provided by the Albert Einstein Center for Fundamental Physics, Bern, Switzerland. For the purpose of open access, the authors have applied a Creative Commons Attribution (CC BY) licence to any Author Accepted Manuscript version arising from this submission. We also acknowledge the contributions of technical and scientific staff to the design, construction, and operation of the MicroBooNE detector as well as the contributions of past collaborators to the development of MicroBooNE analyses, without whom this work would not have been possible.

We particularly acknowledge the many contributions of Salvatore Davide Porzio to devel-
oping this research direction on MicroBooNE.

\end{document}